\documentclass[final,3p,times,twocolumn,sort&compress,longbibliography]{elsarticle}
\newcounter{bla}

\journal{Computer Physics Communications}

\usepackage{color}
\usepackage{hyperref}
\usepackage{amsmath, amssymb}
\hypersetup{colorlinks=true,breaklinks,linkcolor=blue,urlcolor=blue,citecolor=blue}
\usepackage{graphicx}
\usepackage{dcolumn}
\usepackage{bm}
\usepackage{xspace}
\usepackage{physics}
\usepackage{cprotect}

\def\vec#1{\boldsymbol #1}

\newcommand{\HPhi}{\mathcal{H}\Phi}

\def\vec#1{\boldsymbol #1}
\usepackage{natbib}

\usepackage{scalerel}
\usepackage{tikz}
\usetikzlibrary{svg.path}

\definecolor{orcidlogocol}{HTML}{A6CE39}
\tikzset{
  orcidlogo/.pic={
    \fill[orcidlogocol] svg{M256,128c0,70.7-57.3,128-128,128C57.3,256,0,198.7,0,128C0,57.3,57.3,0,128,0C198.7,0,256,57.3,256,128z};
    \fill[white] svg{M86.3,186.2H70.9V79.1h15.4v48.4V186.2z}
                 svg{M108.9,79.1h41.6c39.6,0,57,28.3,57,53.6c0,27.5-21.5,53.6-56.8,53.6h-41.8V79.1z M124.3,172.4h24.5c34.9,0,42.9-26.5,42.9-39.7c0-21.5-13.7-39.7-43.7-39.7h-23.7V172.4z}
                 svg{M88.7,56.8c0,5.5-4.5,10.1-10.1,10.1c-5.6,0-10.1-4.6-10.1-10.1c0-5.6,4.5-10.1,10.1-10.1C84.2,46.7,88.7,51.3,88.7,56.8z};
  }
}

\newcommand\orcidicon[1]{\href{https://orcid.org/#1}{\mbox{\scalerel*{
\begin{tikzpicture}[yscale=-1,transform shape]
\pic{orcidlogo};
\end{tikzpicture}
}{|}}}}

\begin{document}
\begin{frontmatter}



\title{Update of $\mathcal{H}\Phi$: Newly added functions and methods in versions 2 and 3}


\author[issp]{Kota Ido\corref{author}}
\author[itc]{Mitsuaki Kawamura}
\author[issp]{Yuichi Motoyama}
\author[issp]{Kazuyoshi Yoshimi}
\author[nims]{Youhei Yamaji}
\author[ut_phys,ut_int,issp]{Synge Todo}
\author[issp]{Naoki Kawashima}
\author[baqis,issp]{Takahiro Misawa\corref{author}}

\cortext[author] {Corresponding authors.\\\textit{E-mail address:} ido@issp.u-tokyo.ac.jp
\\\textit{E-mail address:} tmisawa@issp.u-tokyo.ac.jp}
\address[issp]{Institute for Solid State Physics, The University of Tokyo,~5-1-5 Kashiwanoha, Kashiwa, Chiba 277-8581, Japan}
\address[itc]{Information Technology Center, The University of Tokyo, Tokyo 113-8658, Japan}
\address[nims]{Research Center for Materials Nanoarchitectonics, National Institute for Materials Science, Namiki, Tsukuba-shi, Ibaraki 305-0044, Japan}
\address[ut_phys]{Department of Physics, The University of Tokyo, Tokyo 113-0033, Japan}
\address[ut_int]{Institute for Physics of Intelligence, The University of Tokyo, Tokyo 113-0033, Japan}
\address[baqis]{Beijing Academy of Quantum Information Sciences, Haidian District, Beijing 100193, China}

\begin{abstract}
$\HPhi$~[{\it aitch-phi}] is an open-source software package
of numerically exact and stochastic calculations for a wide range
of quantum many-body systems. In this paper, we present
the newly added functions and the implemented methods in vers.\,2 and 3.
In ver.\,2, we implement spectrum calculations
by the shifted Krylov method, and low-energy excited 
state calculations by the locally optimal blocking preconditioned
conjugate gradient (LOBPCG) method.
In ver.\,3, we implement the full diagonalization method using ScaLAPACK
and GPGPU computing via MAGMA. 
We also implement a real-time evolution method and 
the canonical thermal pure quantum (cTPQ) state 
method for finite-temperature calculations. The Wannier90 format
for specifying the Hamiltonians is also implemented. 
Using the Wannier90 format, 
it is possible to perform the calculations for the $ab$ $initio$ low-energy
effective Hamiltonians of solids obtained by the open-source software RESPACK. 
We also update Standard mode---simplified input format in $\HPhi$---to use these functions and methods.
We explain the basics of the implemented methods and how to use them. 
\end{abstract}


\end{frontmatter}



{\bf PROGRAM SUMMARY/NEW VERSION PROGRAM SUMMARY}

\begin{small}
\noindent
{\em Program Title:} $\HPhi$~[{\it aitch-phi}]                             \\
{\em CPC Library link to program files:} (to be added by Technical Editor) \\
{\em Developer's repository link:} https://github.com/issp-center-dev/HPhi \\
{\em Code Ocean capsule:} (to be added by Technical Editor)\\
{\em Licensing provisions:} GPLv3  \\
{\em Programming language:} C, Fortran                        \\
{\em Supplementary material:}                                 \\
{\em Journal reference of previous version:} https://www.sciencedirect.com/science/article/\\
pii/S0010465517301200 \\
{\em Does the new version supersede the previous version?:} 
Yes. The latest version has compatibility with the old versions. Although most functions are available in newer versions, some redundant/unnecessary functions were abolished in newer versions.
\\
{\em Reasons for the new version:} Implementation of new 
functions and methods, development of utilities, and bug fixes.\\
{\em Summary of revisions:} We added new functions to obtain excited spectrum, low-energy excited states, and time-dependent physical quantities.
We also implemented the full diagonalization method by using MAGMA on GPGPUs and finite-temperature simulations by using the canonical thermal pure quantum state method.
In addition, we developed utilities for connection with RESPACK and a submodule for a generator of input files.
\\
{\em Nature of problem(approx. 50-250 words):} Physical properties in quantum lattice models with finite system sizes\\
{\em Solution method(approx. 50-250 words):} In $\HPhi$, 
we implemented several numerical methods such as the Lanczos method, the full diagonalization method, the locally optimal block preconditioned conjugate gradient (LOBPCG) method, the real-time evolution method based on the Taylor expansion, the shifted Krylov method, and the microcanonical/canonical thermal pure quantum state method.\\
   \\

\end{small}


\section{Introduction}
To analyze properties of a quantum many-body system, eigenvalues and eigenvectors
of the matrix representations of the Hamiltonians are crucially important.
The exact diagonalization method
\footnote{As we state below, we use the term ``exact diagonalization'' not only for the full diagonalization of a given matrix, but also for the partial diagonalization for obtaining a limited set of the eigenvalues and the eigenvectors.}
is one of the most basic tools for numerically obtaining the eigenvalues and eigenvectors of the Hamiltonians
without any approximations~\cite{Dagotto_RMP1994}.
Since the dimensions of the matrices increase exponentially 
as a function of number of sites $N_{\rm s}$ (for spin-1/2 systems, the number of dimensions 
is given by $2^{N_{\rm s}}$),
{the largest system size handled by the exact diagonalization method is
significantly smaller than that 
of other numerical approaches and severely limited by the available memory size.} 
Nevertheless, numerically exact results for small-size clusters
are useful for
{examining the emergence}
of exotic quantum phases such as quantum spin liquids~\cite{Diep,Balents_Nature2010,Motome_JPSJ2020}
. 
The numerically exact results are also useful for ensuring the accuracy of newly developed methods.

Numerical libraries for the diagonalization of 
the matrices such as LAPACK/ScaLAPACK~\cite{lapack,scalapack}
have been developed and widely used.
These libraries 
support the full and partial diagonalization
of the given matrices stored in the memory, and
the tractable matrix size {for these libraries} is on {the} order of $10^{5}$ at most.
For larger matrices, the power method (or more sophisticated Krylov subspace methods, such as the Lanczos method) 
is a widely used algorithm
for obtaining the {lowest eigenstate (ground state) or
low-lying eigenstates (low-energy excited states)}. 
Since the power method {and the Krylov subspace method mainly consist of}
matrix-vector products, {namely,} 
multiplications of
Hamiltonian matrices by
vectors~\footnote{We often call vectors wave functions.},
the numerical costs and the required memory 
are largely reduced compared to full diagonalization.
Using the Lanczos method,
it becomes feasible in terms of cost to 
obtain the ground state of
36-site spin-1/2 systems, which has a dimensionality of $2^{36}\sim 6.8\times10^{10}${, for example}. 
Moreover, it has been shown
that numerically exact finite-temperature {calculations
are} possible using {a typical pure state}~\cite{Imada_JPSJ1986,Hams_PRE2000,Sugiura_PRL2012,Sugiura_PRL2013,Lloyd,Jin_JPSJ2021},
which can be generated by the power method.
This method is often called the thermal pure quantum (TPQ) state method~\cite{Sugiura_PRL2012,Sugiura_PRL2013}
and can be applied to a wide range of quantum many-body systems. 

In the field of condensed matter physics, 
several pioneering software packages {in which}
the Lanczos method is implemented have been developed,
such as TITPACK~\cite{titpack} for spin-1/2 quantum spin systems,
KOBEPACK~\cite{kobepack} for spin-1 quantum spin systems, 
and SPINPACK~\cite{spinpack} for 
both quantum spin systems and itinerant electron systems.
In {the} {ALPS project~\cite{alet2005alps,albuquerque2007alps,bauer2011alps}}, the exact diagonalization method for a wide
range of the quantum many-body systems is also implemented.
These software packages
have widely been used
for analyzing 
ground states of the quantum many-body systems.
However, inter-process distributed-memory
parallelization is not supported in these software packages. 
Since the number of available cores rapidly increases in modern supercomputers,
the implementation of the hybrid parallelization, i.e., combination
of intra- and inter-node process parallelization, is 
necessary for efficient calculations.

Under those circumstances
we developed and released $\HPhi$ ver.\,1.0 in 2016~\cite{Kawamura_CPC2017,hphi},
which supports the hybrid parallelization combining 
{OpenMP (intra-process shared-memory) and MPI (inter-process distributed-memory)} parallelization.
In $\HPhi$ ver.\,1.0,
we implemented the Lanczos method for obtaining the ground states, the TPQ state method for finite-temperature calculations, 
and the full diagonalization using {the} LAPACK library.
$\HPhi$ has been used to analyze a wide range of quantum many-body systems, 
such as quantum spin liquids
~\cite{Misawa_JPSJ2018,Samarakoon_PRB2018,catuneanu2018path,Ido_PRB2020,Misawa_PRB2020,Xu_PRL2020,Patri_PRR2020,Patri_PRR2020_2,Yamada_PRB2020,Yoshimi_PRR2021,Yoshitake_PRB2020,Misawa_PRR2020,Jang_PRR2020,Laurell_npjQM2020,Jang_PRB2021,Nomura_PRX2021,sala_ncom2021,Yao_PRB2022,Hosoi_PRL2022,Samarakoon_PRR2022,Rayyan_PRB2023}, 
high-$T_{c}$ superconductors~\cite{Zhang_PRR2020,Betto_PRB2021,Iwano_JPSJ2022}, and
correlated topological phases~\cite{Araki_PRR2020,Markov_PRB2021}.
$\HPhi$ is also used to 
examine the accuracy of developed numerical/theoretical methods~\cite{Ido_PRB2020_2,Charlebois_PRX2020,Martinazzo_PNAS2020,Ronto_SciRep2021,Nomura_JPC2021,Nomura_PRL2021,Inui_PRR2021,Dobrautz_PRB2022}.

After the release of $\HPhi$ ver.\,1.0,
we continued to extend the applicable
range of $\HPhi$ by implementing several functions and methods.
Here, we summarize 
the most important functions and methods among the new additions
in
{chronological order}:
\begin{itemize}
\item {Functions} {implementing} spectrum calculation using the shifted Krylov method and the continued-fraction expansion method~\cite{Hoshi_CPC2021,Frommer}~{\bf[ver.\,2.0.0]}
\item {Function implementing} the locally optimal block preconditioned conjugate gradient (LOBPCG) method~\cite{Knyazev_SIAM2001} for low-energy excited state calculations~{\bf[ver.\,2.0.0]}
\item {Function implementing} real-time evolution~{\bf[ver.\,3.0.0]}
\item {Function implementing} full diagonalization using ScaLAPACK~\cite{scalapack}~{\bf[ver.\,3.1.0]}
\item {Function implementing} full diagonalization using GPGPU via MAGMA~\cite{MAGMA}~{\bf[ver.\,3.1.0]}
\item {Interface to} RESPACK~\cite{Nakamura_CPC2021,respack}, which derives $ab$ $initio$ effective Hamiltonians of solids~{\bf[ver.\,3.3.0]}
\item Making {{\it Standard mode}} submodule (Standard mode will be explained in Section \ref{sec:standard})~{\bf[ver.\,3.4.0]}
\item {Functions implementing} the canonical TPQ (cTPQ) method~\cite{Sugiura_PRL2013}~{\bf[ver.\,3.5.0]}
\end{itemize}
In the following sections of the present paper,
we explain the basics of these functions and how to use them.
We also show 
several examples of the implemented methods. 
Note that we do not explain the basic usage of  $\HPhi$  and the methods/functions implemented in $\HPhi$ ver.\,1 in this paper since they are already explained
in Ref. \cite{Kawamura_CPC2017}. In this paper, we mainly explain the above newly added functions and methods. For the basic usage of $\HPhi$, please see the manual~~\cite{HPhi_manual}
as well as the tutorials~\cite{HPhi_tutorial}. 

The present paper is organized as follows: 
In Section \ref{Sec:Installation}, we explain
how to install $\HPhi$ and the submodule repository of Standard mode.
In Section \ref{sec:fulldiag},
we explain full diagonalization with ScaLAPACK~\cite{scalapack} on multiple CPUs and with MAGMA on multiple GPGPUs~\cite{MAGMA}.
We also explain
{how to output matrices generated by the $\HPhi$ routines
and how to input matrices prepared by the users}
in the Matrix Market {exchange format for general sparse matrices}.
In Section \ref{sec:spectrum}, we explain spectrum
calculation using $\HPhi$. In $\HPhi$, in addition to the
conventional continued-fraction expansion method~\cite{Dagotto_RMP1994}, the shifted-Krylov method~\cite{Frommer,Yamamoto_JPSJ2008}
is implemented. In Section \ref{sec:CG}, we detail the LOBPCG method~\cite{Knyazev_SIAM2001,Yamada_JSCES2006}, which
enables us to obtain several low-energy excited states {simultaneously}.
We also show the effects of the preconditioning in the LOBPCG method.
In Section \ref{sec:timeevo}, we
{explain} how to perform 
real-time evolution using $\HPhi$. 
In Section \ref{sec:cTPQ}, we review the basics of the
TPQ method. 
We also show that the cTPQ method
{indeed} reproduces the results
of the full diagonalization for the one-dimensional Heisenberg chain.
In {Section} \ref{sec:standard}, we summarize the methods, models, and lattices
available in Standard mode. 
In {Section} \ref{sec:wan90}, we give {an} explanation of
the interface to the Wannier90 format~\cite{wan90_HP,Pizzi_2020JPC} of the Hamiltonians. 
In {Section} \ref{sec:respack},
we explain the connection between {$\HPhi$ and} RESPACK~\cite{Nakamura_CPC2021,respack},
which derives the $ab$ $initio$ low-energy effective Hamiltonians for solids from the outputs of the plane-wave density functional theory codes.
Finally, Section \ref{sec:summary} is devoted to the summary and future development of $\HPhi$.

\section{Installation}
\label{Sec:Installation}
Before explaining the newly added functions,
we explain how to download and install $\HPhi$.
We also explain the submodule repository of Standard mode.

\subsection{How to download and install $\HPhi$}
$\HPhi$ can be downloaded through the following GitHub repository:
\begin{verbatim}
https://github.com/issp-center-dev/HPhi
\end{verbatim}
The latest version of $\HPhi$ is ver.\,3.5.1.
One can check the history and brief summaries of 
the releases on the release page~\cite{HPhi_release}.
On this page, 
one can download the gzipped tar files for each version, 
which contain source codes and manuals.

After ver.\,3.0.0, we ended support for
\verb|HPhiconfig.sh|, which generates Makefiles.
In the current version, to build $\HPhi$, 
it is necessary to use the CMake utility~\cite{cmake} as follows:
\begin{verbatim}
$ cmake -B build -DCONFIG=$Config
$ cmake --build build
\end{verbatim}
Typical examples of \verb|$Config| are given as follows:
\begin{itemize}
\item \verb|gcc| : GCC
\item \verb|intel|: intel compiler + MKL library
\end{itemize}

To check whether $\HPhi$ has been installed correctly, we add several tests for CTest utility. One can run Ctest utility in the build directory by the following command.
\begin{verbatim}
$ cmake --build build --target test
\end{verbatim}

To use ScaLAPACK for full diagonalization, 
it is necessary to specify the cmake option as
\begin{verbatim}
cmake -DUSE_SCALAPACK=ON
\end{verbatim}
As we will explain in the next section, 
$\HPhi$ also supports full diagonalization by GPGPU via MAGMA.
To use this function, it is necessary to install MAGMA.
For detailed information on MAGMA including how to install it, 
one can refer to the official web page~\cite{MAGMA_install}.

\subsection{Submodule repository of Standard mode}

In Standard mode of $\HPhi$, one can conduct the calculations for predefined systems
by preparing one simple and human-readable input file. 
Using Standard mode, the users can also generates all the necessary input files for Expert mode from one input file for Standard mode.
For details of Standard and Expert modes, please refer to 
Ref. \cite{Kawamura_CPC2017} and
the manual~\cite{HPhi_manual}.
Standard mode is also used in \verb|mVMC|~\cite{Misawa_CPC2019,mVMC}, which is a software package for the many-variable variational Monte Carlo method~\cite{Tahara_JPSJ2008}. 
These two packages share the main part of Standard mode.
Therefore, to make the maintenance of Standard mode easier,
we separate Standard mode as a submodule in the independent repository at
\begin{verbatim}
https://github.com/issp-center-dev/StdFace
\end{verbatim}
By using the following command, 
users can use the updated Standard mode in both $\HPhi$ and mVMC:
\begin{verbatim}
$ git submodule update -r -i
\end{verbatim}

\section{Newly added methods/functions}
\label{Sec:Methods}
\subsection{Full diagonalization using GPGPU and ScaLAPACK}
\label{sec:fulldiag}
In $\HPhi$ ver.\,1.0, full diagonalization was only implemented by {calling} 
\verb|zheev|
{routine} in LAPACK, which only supports the 
{OpenMP} parallelization in {a} single CPU processor. 
In $\HPhi$ ver.\,3.1.0, 
we implement full diagonalization by ScaLAPACK,
which enables us to perform full diagonalization by inter-process {distributed-memory} parallelization.
It is 
necessary to
specify {the keyword} \verb|Scalapack| as {``}\verb|Scalapack=1|{"} in the input file for {Standard} mode{,
or add a line} {``}\verb|Scalapack 1|{"} in \verb|calcmod.def| for {\it Expert mode} 
(Details of Expert mode are explained in Ref.~\cite{Kawamura_CPC2017}).
Only by either one of these two minor modifications to the input file,
one can perform
full diagonalization by {calling} {Sca}LAPACK {routine}. 
The cost of the full diagonalization for matrices of $O(10^5)$ linear dimensions becomes
reasonable/cheap by ScaLAPACK.

We also implement full diagonalization {on} GPGPUs by {calling a MAGMA routine} \verb|magma_zheevd|~\cite{MAGMA}.
If MAGMA is installed on the user's environment,
by specifying the number of GPGPUs (e.g., \verb|NGPU=2| in the input file of {Standard} mode or 
\verb|NGPU 2| in \verb|calmod.def| for {Expert} mode),
one can perform full diagonalization {on multi-GPGPU environments}. 

\subsubsection{Performance of the full diagonalization}
\begin{figure}[t] 
\begin{center} 
\includegraphics[width=0.48\textwidth]{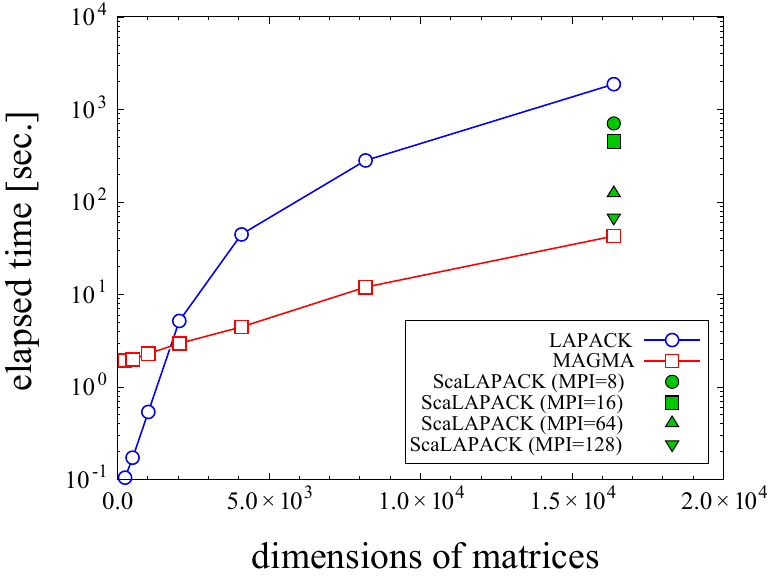}
\caption{
Comparison of elapsed time for full diagonalization
for the one-dimensional antiferromagnetic Heisenberg chain by
the {\bf zheev} routine in LAPACK with 16 {OpenMP} threads, and
the {\bf magma\_zheevd} routine with 16 OpenMP threads and a single GPGPU, NVIDIA HGX A100 40GB.
The calculations are conducted at the supercomputer system C (kugui)
{at the ISSP Supercomputer Center, the University of Tokyo}.
The calculations are done {for several system sizes,
from $L=8$ ({the} matrix dimensions are $2^8=256$)
to $L=14$ ({the} matrix dimensions are $2^{14}=16384$).}
We also show the results obtained with ScaLAPACK with 16 {OpenMP} threads and $L=14$.
In ScaLAPACK, by increasing the number of {the MPI processes}, 
the elapsed time becomes shorter.
However, full diagonalization {on a single} GPGPU is
still faster than using ScaLAPACK with 128 MPI processes.}
\label{fulled}
\end{center}
\end{figure}

To benchmark the different full diagonalization methods,
we performed full diagonalization for the one-dimensional Heisenberg model.
An example of the input file is given as follows:
\begin{verbatim}
L       = 12
model   = "SpinGC"
lattice = "chain"
method  = "FullDiag"
J       = 1.0
\end{verbatim}
By changing the system size $L$,
the dimensions of the target matrices $d_{\rm H}=2^{L}$ are increased.
In Fig.~\ref{fulled},
we show the elapsed time of the full diagonalization as a function of $d_{\rm H}$ for $L=8$--$14$.
For $d_{\rm H}\geq 2500$, the full diagonalization by {a single} GPGPU is faster than
that by
a single CPU with 16 cores.
For $d_{\rm H}=2^{14}=16384$, the full diagonalization {on}
GPGPU {by the MAGMA routine} is about ten times faster.
We also find that the full diagonalization {on a single GPGPU} by MAGMA is still faster than that by
ScaLAPACK using 16 nodes ($2048=128\times16$ cores).
This result demonstrates that GPGPU computing is an efficient tool for performing full diagonalization 
of matrices whose dimensions are 
around $d_{\rm H}\sim 10^{4}$. 

We note that the elapsed time summarized in Fig.~\ref{fulled} does not include that for calculations of physical quantities.
If the full diagonalization method has been performed by using Standard mode, the elapsed time for
calculations of physical quantities
will be significantly longer than that of the full diagonalization of the Hamiltonian matrix. 
When we specify \verb|OneBodyG| and \verb|TwoBodyG| in \verb|namelist.def|, 
correlation functions {or static Green's functions} defined in \verb|OneBodyG| and/or \verb|TwoBodyG| are calculated for all the eigenstates.
In {these calculations on multi-process parallel environments},
the eigenvectors in each MPI process are gathered into
a single process and then the correlation functions are calculated in that process.
This communication process becomes the most time-consuming part of the whole simulation when we use ScaLAPACK with a large number of MPI processes.
For example, compared with the elapsed time of the full diagonalization by ScaLAPACK with 128 MPI processes
summarized in Fig.\,\ref{fulled}, 
it takes about ten times longer to calculate {the} correlation functions generated by {Standard} mode.

\subsubsection{Output/Input by Matrix Market exchange format}
In $\mathcal{H}\Phi$, it is also possible to 
output the {Hamiltonian matrices} of the quantum many-body systems
in the Matrix Market exchange format~\cite{MatrixMarket}
by specifying the keyword \verb|HamIO="out"|
in Standard mode or \verb|OutputHam=1| in
\verb|calcmode.def| in Expert mode.
Here, we show {an} example of the output file 
(\verb|zvo_Ham.dat|) in the
Matrix Market exchange {format}:
{A} simple $2\times 2$ matrix,
{\begin{align}
\mathcal{H}=
\begin{pmatrix}
1 & -i \\
i & -1
\end{pmatrix},
\end{align}
is outputted as follows:}
\begin{verbatim}
% 2by2
2 2 4 
1 1 1.0  0.0
1 2 0.0  -1.0
2 1 0.0 1.0
2 2 -1.0  0.0
\end{verbatim}
In the first line, we write the
dimensions of the target matrix ({the} number of columns is $2$ {and the} number of rows is $2$) and the number of non-zero elements (in this case, $4$).
Other than the first line,
{the row (column) index, $\ell$ ($m$), and the real (imaginary) part of the matrix element, $u$ ($v$), 
are written in the order of} 
\begin{align}
{\ell}~~~{m}~~~u~~~v {.}
\end{align}
We note that indices of the row and column start from one (a start offset of one) as in the standard
notation of Fortran.
This function is useful for examining the structure of the matrices
of quantum many-body systems. 
By specifying keywords in {\tt modpara.def} and {\tt calcmod.def}, files in the Matrix Market exchange {format}
can also be used as inputs for calculations of the full diagonalization method.
The details are explained in $\HPhi$'s manual~\cite{HPhi_manual}. 

\subsection{Spectrum calculation} 
\label{sec:spectrum}
Using the eigenvectors, $\HPhi$ can calculate the 
dynamical correlation {function}, which is defined as
\begin{align}
D(z)&=\ev{\qty[z I-\hat{H}]^{-1}}{\Phi_{\rm ex}},\\
\ket{\Phi_{\rm ex}}&=\hat{O}\ket{\Phi_{0}},
\end{align}
where $z$ is a given complex number, $I$ is an identity matrix, $\hat{H}$ is a Hamiltonian matrix, $\ket{\Phi_{0}}$ is the target eigenvector, and
$\hat{O}$ is the excitation operator. 
As {excitation} operators, $\HPhi$ implements
single-particle excitations (\verb|SingleExcitation|) and pair excitations (\verb|PairExcitation|)
defined as
\begin{align}
&\hat{O}_{\rm single}=
\begin{cases}
&\sum_{i,\sigma}A_{i\sigma}c_{i\sigma}^{\dagger}\\
&\sum_{i,\sigma}A_{i\sigma}c_{i\sigma}
\end{cases}{,}
\\
&\hat{O}_{\rm pair}=
\begin{cases}
&\sum_{i,j,\sigma,\tau}A_{i\sigma,j\tau}c_{i\sigma}^{\dagger}c_{j\tau}\\
&\sum_{i,j,\sigma,\tau}A_{i\sigma,j\tau}c_{i\sigma}c^{\dagger}_{j\tau}
\end{cases}{,}
\end{align}
{where $c_{i\sigma}^{\dagger}$ ($c_{i\sigma}$) represents
a creation (annihilation) operator in the second quantization, which
generates (removes) a particle with $\sigma$ spin at the $i$th site/orbital.}
We note that the users can specify the order of operators 
(order of the creation and annihilation operators) in the input file.

$\HPhi$ implements two different algorithms 
for calculating the
dynamical correlation function $D(\omega)$. 
One is the continued-fraction expansion based on the Lanczos method~(\verb|method="Lanczos"|)~\cite{Gagliano_PRL1987,Dagotto_RMP1994}
which is frequently used in previous studies.
Another is the shifted Krylov-subspace method~(\verb|method="CG"|)~\cite{Frommer,Yamamoto_JPSJ2008}.
Since the details of {the implementation of the} shifted Krylov-subspace method are
already {given} in Ref.~\cite{Hoshi_CPC2021}, 
we just note the advantages of the shifted Krylov-subspace method
{in comparison} with the conventional continued-fraction expansion. 
The advantages of the shifted Krylov-subspace method are
summarized as follows:
\begin{enumerate}
\item We can explicitly evaluate the convergence of $D(z)$.
\item We can calculate the off-diagonal dynamical correlation functions, i.e.,
$D_{ab}(z)=\mel*{\Phi_{0}}{\hat{O}_{a}^{\dagger}\qty[z I-\hat{H}]^{-1}\hat{O}_{b}}{\Phi_{0}}$.
\end{enumerate}
On the first point, 2-norms
of the residual vectors are output in \verb|residual.dat| 
The definition of the residual vector in the shifted Krylov {subspace} method
is detailed in Ref.~\cite{Hoshi_CPC2021}.
By examining \verb|residual.dat|, one can {quantify}
the accuracy of the convergence.
On the second point, although $\HPhi$ does not support the function of computing
off-diagonal dynamical correlation functions {($a\ne b$)}, it is possible 
to calculate them by the shifted Krylov subspace method.

\begin{figure}[t] 
\begin{center} 
\includegraphics[width=0.48 \textwidth]{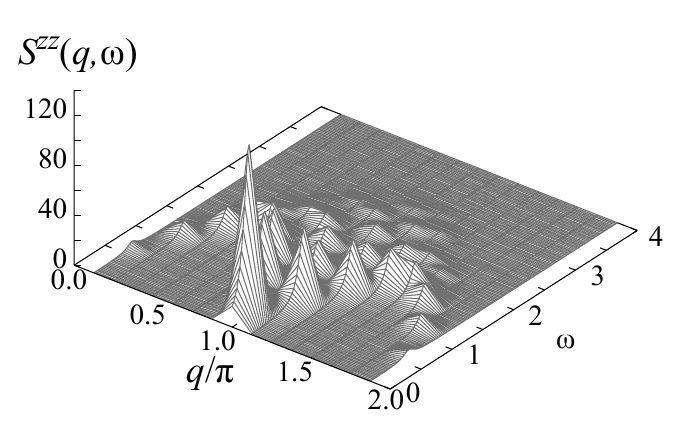}
\caption{Dynamical spin structure factors
for the one-dimensional antiferromagnetic Heisenberg chain. 
We take $\eta=0.1$.
We can see the signature of the
des Cloizeaux-Pearson
mode~\cite{Cloizeaux_PR1962}. 
}
\label{Sqw}
\end{center}
\end{figure}

As an example of the dynamical correlation functions,
we calculate the dynamical spin structure factors for the
one-dimensional antiferromagnetic Heisenberg model defined as
\begin{align}
&S^{zz}(q,\omega)=-{\rm Im}\qty[\ev{\qty[(\omega+i\eta)I-\hat{H}]^{-1}}{\Phi_{\rm ex}(q)}],\\
&\ket{\Phi_{\rm ex}}=\hat{O}_{\rm spin}\ket{\Phi_{\rm GS}},\\
&{\hat{O}_{\rm spin}=\sum_{\ell=0}^{L-1}\hat{S}_{\ell}^{z}e^{iq\ell}\ket{\Phi_{\rm GS}}},
\end{align}
where $\ket{\Phi_{\rm GS}}$ is the ground state{, $\hat{S}_{\ell}^z = (c^{\dagger}_{\ell\uparrow}c_{\ell\uparrow}
-c^{\dagger}_{\ell\downarrow}c_{\ell\downarrow})/2$,}
and $q$ $[=2\pi m/L\ (m \in \mathbb{Z})]$ represents
the wave numbers.
In Fig.~\ref{Sqw}, we show $S_{zz}(q,\omega)$ for the $L=16$ Heisenberg chain 
obtained by the shifted Krylov method.
{The calculation flow chart} is summarized as follows:
\begin{enumerate}
\item Calculating the ground state $\ket{\Phi_{\rm GS}}$
\item Preparing the \verb|pair.def| that denotes {$\hat{O}_{\rm spin}=\sum_{\ell}\hat{S}_{\ell}^{z}e^{iq\ell}$} for each $q$
\item Calculating $S^{zz}(q,\omega)$ using the shifted-Krylov method \verb|method="CG"| in \verb|calmod.def|
\end{enumerate}
These three steps can be performed automatically
by using a script
\verb|Spectrum.py| in Ref.~\cite{repo}.
By executing {the} following command,
one can calculate $S^{zz}(q,\omega)$ for {the} $16$-site one-dimensional
Heisenberg chain{:}
\begin{verbatim}
$ python3 Spectrum.py 16
\end{verbatim}

\subsection{Iterative diagonalization method for multiple eigenvectors}
\label{sec:CG}
The locally optimal block preconditioned conjugate gradient (LOBPCG) method~\cite{Knyazev_SIAM2001} is 
implemented in $\HPhi$. Using the LOBPCG method,
we can simultaneously obtain several low-energy excited states in one calculation.
Since the orthogonality of each eigenvector is automatically guaranteed,
we can also determine the degeneracy of the low-energy excited states.
Although the Lanczos method can also be used to obtain the low-energy excited states, 
it needs additional treatment, such as the re-orthogonalization of the obtained eigenvectors. 
As we detail below, the procedure of the LOBPCG method is simple and straightforward.

\begin{figure}[t] 
    \begin{center} 
    \includegraphics[width=0.48 \textwidth]{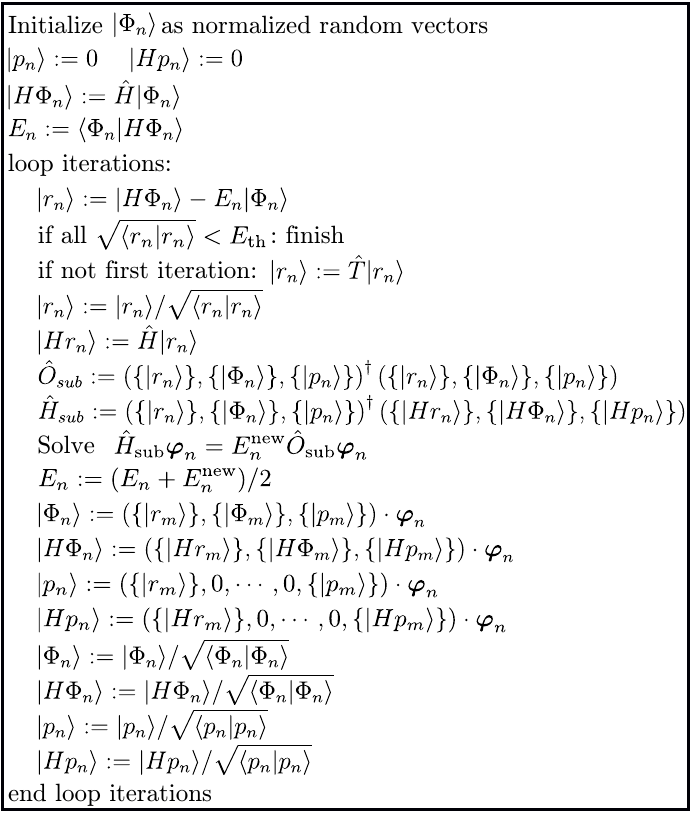}
    \caption{Algorithm of LOBPCG method. Here, $n$ is the index of eigenstates and $E_{th}$ is the 
        energy threshold for the convergence criteria. $\{\cdots\}$ indicates a set whose size is the number of the target eigenstates.
    }
    \label{lobpcg_alg}
   \end{center}
\end{figure}

Here, we briefly explain the algorithm of the LOBPCG method.
In the LOBPCG method, in addition to the candidate vectors of the eigenstates $\{\ket{\Phi_n}\}$, 
we calculate residual vectors $\{\ket{r_n}\}$, 
conjugate gradient vectors $\{\ket{p_n}\}$, 
and those to which the Hamiltonian is applied $\{\ket{H \Phi_n}\}$, $\{\ket{H r_n}\}$, $\{\ket{H p_n}\}$ 
($n=0,\cdots,M-1$, where $M$ is the number of the target eigenstates).
Using these vectors, we define the following
Hamiltonian in the subspace and the overlap matrix as
\begin{align}
  &\hat{H}_{\rm sub}=\ev*{\hat{H}}{V},\\
  &\hat{O}_{\rm sub}=\braket{V},\\
  &\ket{V}=\qty(\{\ket{r_n}\},\{\ket{\Phi_n}\},\{\ket{p_n}\}).
\end{align}
We note that the dimension of {$\hat{H}_{\rm sub}$} is the number of target eigenstates times three, namely $(3M)$.
We solve the following generalized eigenvalue equation for the subspace:
\begin{align}
  &\hat{H}_{\rm sub} \vec{\varphi}_n = E_n^{\rm new} \hat{O}_{\rm sub} \vec{\varphi}_n.
\end{align}
In the actual calculation, we solve the equation using LAPACK.
By using obtained eigenvectors $\vec{\varphi}_{n}$,
we update $\{\ket{\Phi_{n}}\}$,$\{\ket{H\Phi_n}\}$,$\{\ket{p_n}\}$, and $\{\ket{Hp_n}\}$  as follows:
\begin{align}
\ket{\Phi_{n}}&=(\{\ket{r_m}\},\{\ket{\Phi_{m}}\},\{\ket{p_{m}}\})\cdot\vec{\varphi}_{n},\\
\ket{H\Phi_{n}}&=(\{\ket{Hr_m}\},\{\ket{H\Phi_{m}}\},\{H\ket{p_{m}}\})\cdot\vec{\varphi}_{n},\\
\ket{p_{n}}&=(\{\ket{r_m}\},0,\cdots,0,\{\ket{p_{m}}\})\cdot\vec{\varphi}_{n},\\
\ket{Hp_{n}}&=(\{\ket{Hr_m}\},0,\cdots,0,\{\ket{Hp_{m}}\})\cdot\vec{\varphi}_{n},
\end{align}
where we use the simplified notation
\begin{align}
&\ket{\Phi_{n}}=(\{\ket{r_m}\},\{\ket{\Phi_{m}}\},\{\ket{p_{m}}\})\cdot\vec{\varphi}_{n} \\
&=\sum_{m=0}^{M-1}\qty[\varphi_{nm}\ket{r_{m}}+\varphi_{n,m+M}\ket{\Phi_{m}}+\varphi_{n,m+2M}\ket{p_{m}}].
\end{align}
Here, $\varphi_{nm}$ is the $m$th component of the $n$th eigenvector $\vec{\varphi}_{n}$.
Finally, we normalize the updated vectors, 
for example, $\ket{\Phi_{n}} \leftarrow \ket{\Phi_{n}}/|\ket{\Phi_{n}}|$.
By performing the procedure iteratively, we can shift the subspace to the lower energy manifold.
Figure \ref{lobpcg_alg} summarizes the procedure of the LOBPCG method.

In each iteration, we multiply the residual vectors $\ket{r_{n}}$ by the preconditioner $\hat{T}$
to accelerate the convergence. 
It is known that the convergence becomes faster if the eigenvalues
are distributed around 1~\cite{Murota_2020}.
Therefore, the ideal preconditioner is given by $\hat{T}=(\hat{H}-E_n)^{-1}$,
which makes the modified eigenvalues around the 
ground states located around 1 ($\tilde{E}_{0}=E_{0}/E_{0}=1, \tilde{E}_{1}=E_{1}/E_{0}\sim 1,\dots$).
However, it is impossible to take  $\hat{T}=(\hat{H}-E_n)^{-1}$. Thus,
one of the main problems to be addressed in preconditioning
is how to approximate $(\hat{H}-E_n)^{-1}$. 
By taking the approximate form of $(\hat{H}-E_n)^{-1}$ appropriately,
it is possible to accelerate the convergence~\cite{Yamada_JSCES2006}.

\begin{figure}[t] 
    \begin{center} 
    \includegraphics[width=0.28 \textwidth]{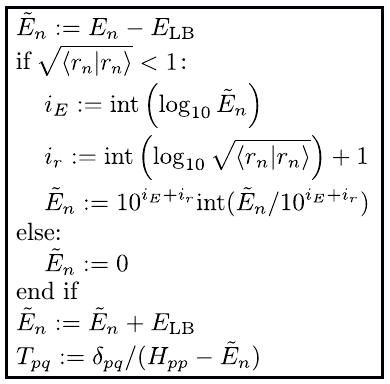}
    \caption{
        Algorithm of the adaptively shifted point Jacob method.
        $E_{\rm LB}$ is a constant that is lower than the ground-state energy $E_{\rm GS}$ ($E_{\rm LB}<E_{\rm GS}$).
    }
    \label{precg}
   \end{center}
\end{figure}

In $\HPhi$, following Ref.~\cite{Yamada_JSCES2006},
we implement the adaptively shifted point Jacob method
to approximate $(\hat{H}-E_n)^{-1}$.
In this method, we only consider the diagonal elements of $(\hat{H}-E_n)^{-1}$
and approximate them as
\begin{align}
&T_{pq}=\frac{\delta_{pq}}{H_{pp}-\tilde{E}_{n}},
\label{eq_pointjacob}
\end{align}
where $\tilde{E}_{n}$ is determined by the procedure
shown in Fig.~\ref{precg}.
Since we adaptively determine $\tilde{E}_{n}$ to approach the true 
eigenvalue from below ($E_{\rm LB}\leq \tilde{E}_{n}<E_{\rm GS}$, where $E_{\rm LB}$ is an estimated lower-bound constant of the ground-state energy and $E_{\rm GS}$ is the true ground-state energy), 
the preconditioning increases the weights of 
eigenvectors around the ground states. 
This indicates that the preconditioning 
can efficiently extract the weights of the eigenvectors around the ground states.
In contrast to this, if we use ${E}_n$, which is always
larger than the true ground-state energy, 
the preconditioning increases the weights of
eigenvectors other than the ground states.
It is pointed out that the use of $E_{n}$ sometimes causes unstable convergence~\cite{Yamada_JSCES2006}.

The concrete procedure of the preconditioning is shown in Fig. \ref{precg}.
In the point Jacob method, since we just approximate $\hat{T}$ as a diagonal matrix,
we need no extra memory or computational cost. 
Nevertheless, as we show later, the convergence can be accelerated.
In the actual calculations,
the lower bound of the ground-state energy $E_{\rm LB}$ is estimated from 
the parameters of the second quantized Hamiltonian ($t$, $J$, $U$, and so on).
We note that the Gershgorin circle theorem~\cite{Gershgorin} is often used to estimate
the lower bound of the ground-state energy.
However, to use this theorem, it is necessary to evaluate all non-zero matrix elements and its numerical cost is large
for large system sizes.
Thus, we do not use the Gershgorin circle theorem but use a simple estimation of
the lower bound from the parameters in the Hamiltonian.

\begin{figure}[t] 
    \begin{center} 
    \includegraphics[width=0.48 \textwidth]{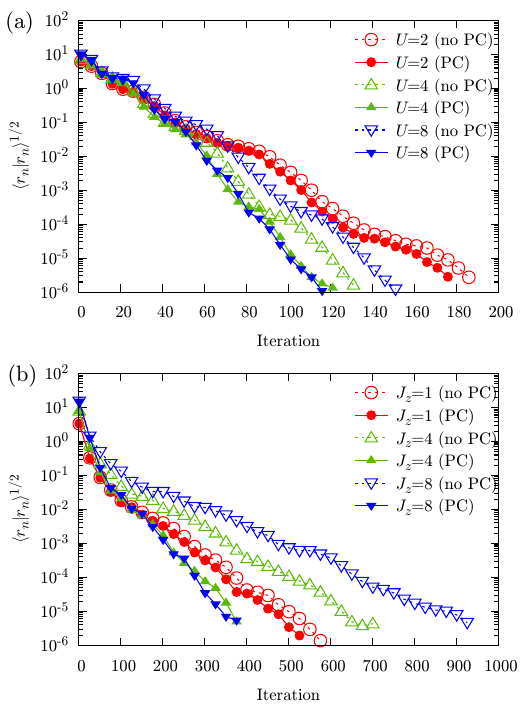}
    \caption{
        Convergence of the LOBPCG method for (a) the 16-site Hubbard model on the square lattice ($t=-1$)
        and (b) the 30-site $XXZ$ model on the kagome lattice ($J_x=J_y=1$).
        Dashed (solid) lines with empty (filled) symbols indicate the result without (with) preconditioning.
        Circles and upward and downward triangles in (a) [(b)] indicate the result of $U$ = 2, 4, and 8 ($J_z$ = 1, 4, and 8), respectively. In this calculation, we obtain 32 eigenstates simultaneously.  
    }
    \label{residual}
   \end{center}
\end{figure}

Here, we examine the effects of the preconditioning.
Figure \ref{residual} shows 
the convergence behavior of the LOBPCG method in the 
16-site Hubbard model on the square lattice and the 
30-site $XXZ$ model on the kagome lattice.
We vary the parameters $U$ and $J_z$ to see the preconditioning 
effects since these parameters affect the diagonal part of the Hamiltonian. 
In all cases, we can confirm that the preconditioning accelerates the convergence.
We can also confirm that the convergence becomes faster
for large $U$ and large $J_{z}$.
This is because the $\hat{T}$ matrix becomes a good approximation 
of $(\hat{H}-E_n)^{-1}$ for larger $U$ or $J_z$ since
the absolute value of the diagonal part increases.
For the standard Hamiltonians such as the Heisenberg model or the Hubbard model,
we can confirm that the preconditioning is efficient.
However, it is not clear how the preconditioning affects the convergence
in general Hamiltonians. Therefore, in the current version of $\HPhi$,
preconditioning is switched off in the default settings, i.e., $\hat{T}=I$.
If users want to examine the effects of preconditioning,
please add the following parameter in \verb|modpara.def|. 
\begin{verbatim}
PreCG 1
\end{verbatim}

We next show another example of an application of the LOBPCG method.
By using the LOBPCG method, we obtain the low-energy excited states of the 
two-dimensional antiferromagnetic Heisenberg model ($N_{\rm s}=4\times4$, total $S_z$ is zero). 
An input file is given as follows.
\begin{verbatim}
L       = 4
W       = 4
lattice = "square"
model   = "Spin"
method  = "CG"
J       = 1
2Sz     = 0
exct    = 32
\end{verbatim}

\begin{figure}[t] 
\begin{center} 
\includegraphics[width=0.48 \textwidth]{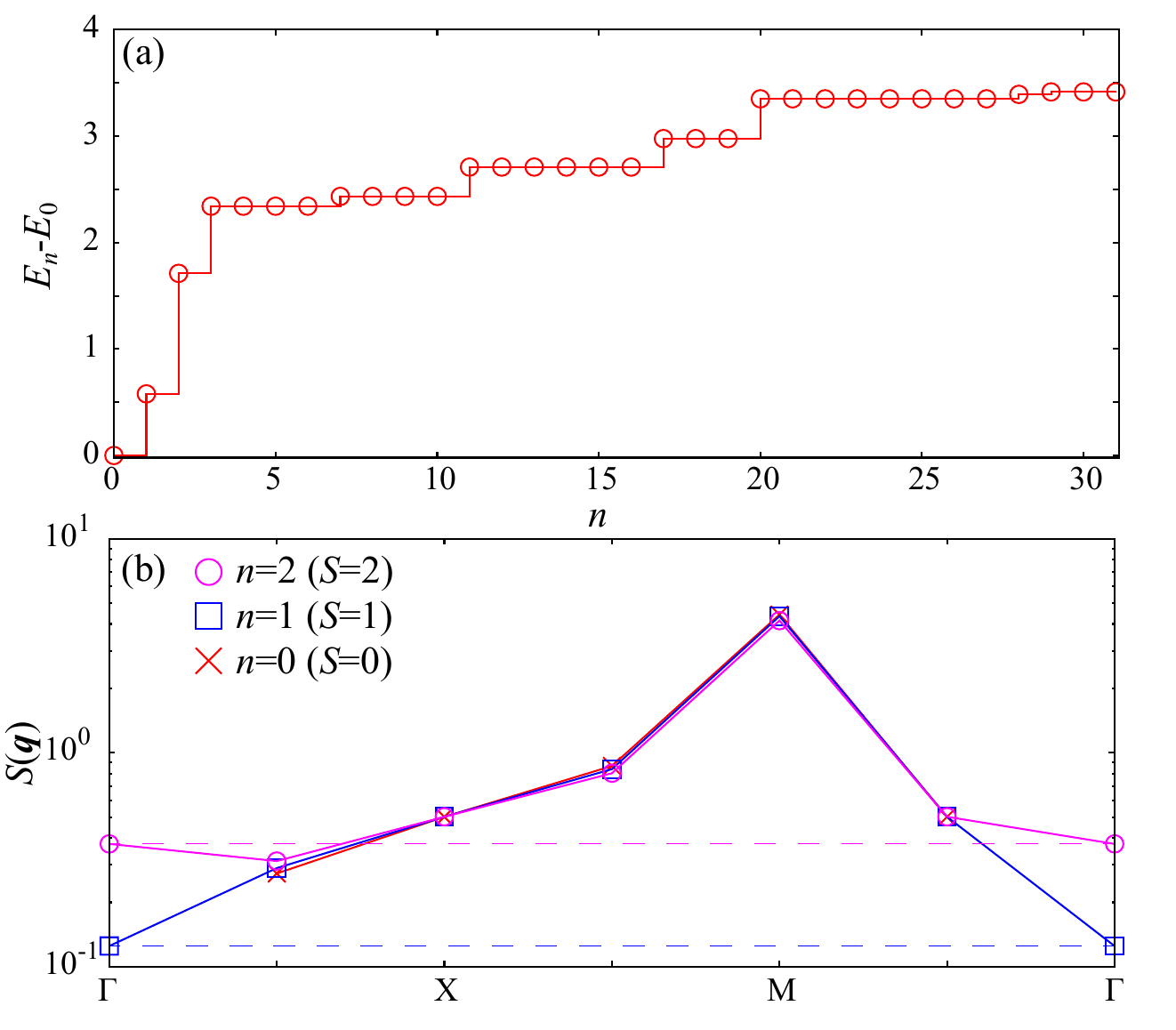}
\caption{
(a) Energies of 32 low-energy excited states measured from
the ground-state energy $E_{0}$ for $4\times4$ antiferromagnetic Heisenberg model (total $S_{z}=0$ sector).
(b) Spin structure factors for three low-energy excited states. 
The spin structure factors at $\vec{q}=0$ are given by $S(\vec{q})=S(S+1)/N_{\rm s}$, where $S$ is the total spin.
The horizontal broken lines show $S(S+1)/N_{\rm s}$ for $S=1$ and $S=2$.
The amplitude of $S(\vec{0})$ for the ground state is less than $10^{-8}$.
}
\label{CG_4by4}
\end{center}
\end{figure}

In Fig.~\ref{CG_4by4}, we show the eigenenergies obtained by the LOBPCG method.
We simultaneously obtain 32 low-energy excited states.
By calculating the spin structure factors, we can see that
the triplet ($S=1$) [the quadruplet ($S=2$)] state is the first [second] excited state.
As is shown in this example, the LOBPCG method offers information on
the quantum numbers and the degeneracies of the low-energy excited states,
which are often used for identifying the nature of quantum magnets.

\subsection{Real-time evolution}
\label{sec:timeevo}
For a given wave function (not limited to an eigenstate),
one can perform real-time evolution by solving the
following time-dependent Schrodinger equation:
\begin{align}
i\frac{\partial\ket{\Phi(t)}}{\partial t}=\hat{H}(t)\ket{\Phi(t)}.
\end{align}
By discretizing time with the width $\Delta t$,
we can obtain the solution of the Schrodinger equation as follows:
\begin{align}
&\ket{\Phi(t+\Delta t)}=\exp[-i\hat{H}\Delta t]\ket{\Phi(t)},\\
&\sim\sum_{n=0}^{n_{\rm max}}\frac{1}{n!}\qty(-i\hat{H}\Delta t)^n\ket{\Phi(t)}.
\end{align}
This indicates that real-time evolution can be done by multiplying
the Hamiltonians by the wave functions.
The default values are set as $n_{\rm max} = 10$ and $\Delta t=0.01$ from  Standard mode ver.\,0.5.
The unit of the time $t$ and the time grid $\Delta t$ is the inverse of that of the energy.
By checking $\braket{\Phi(t)}{\Phi(t)}=1$, one can examine the 
accuracy of the real-time evolution.

To perform real-time evolution,
it is necessary to specify the method as 
\begin{verbatim}
method = "Time-Evolution"
\end{verbatim}
in \verb|calmode.def|.

For real-time evolutions, we provide two modes, Expert mode and Standard mode, 
for specifying time-dependent Hamiltonians. 
In Expert mode, users can define time-dependent Hamiltonians 
with arbitrary one- and two-body terms represented as
\begin{eqnarray}
\hat{H}(t) &=& - \sum_{i,j} t_{ij}(t) c^{\dagger}_{r_i \sigma_i}c_{r_j \sigma_j} \nonumber \\
&+& \sum_{i,j,k,l} I_{ijkl}(t) c^{\dagger}_{r_i \sigma_i} c_{r_j \sigma_j} c^{\dagger}_{r_k \sigma_k} c_{r_l \sigma_l}, 
\end{eqnarray}
where $c^{\dagger}_{r \sigma}$ and $c_{r \sigma}$ denote respectively the creation and annihilation operators of a particle with spin $\sigma$ ($\sigma=\uparrow {\rm or} \downarrow$) at site $r$.
In Standard mode, $\HPhi$ can perform real-time simulations on 
typical nonequilibrium systems such as in quench dynamics and electron systems irradiated by a laser.
For details on specifying the time-dependent Hamiltonians,
see the manual of $\HPhi$~\cite{HPhi_manual}.

As an example of quantum dynamics using $\HPhi$, we show sudden quench dynamics in the one-dimensional Hubbard model at half filling.
The Hamiltonian is defined by the following equation:
\begin{equation}
\hat{H}_{\rm quench}(t) = - t_{\rm hop} \sum_{\ev{i,j}} \sum_{\sigma} c^{\dagger}_{i\sigma}c_{j\sigma} + U(t) \sum_{i}^{N_{\rm s}} n_{i\uparrow} n_{i\downarrow}, \nonumber\\ \label{Hquench} 
\end{equation}
where $\ev{i,j}$ denotes the indices of the nearest neighbor sites $i$ and $j$ on the one-dimensional chain, $N_{\rm s}$ denotes the system size, and 
\begin{align}
&U(t) = \begin{cases}
  U_{0} & (t < 0), \\
  U_{\rm quench} & (t \geq 0). \\
  \end{cases}
\label{Uquench}
\end{align}

\begin{figure}[t] 
\begin{center} 
\includegraphics[width=0.48 \textwidth]{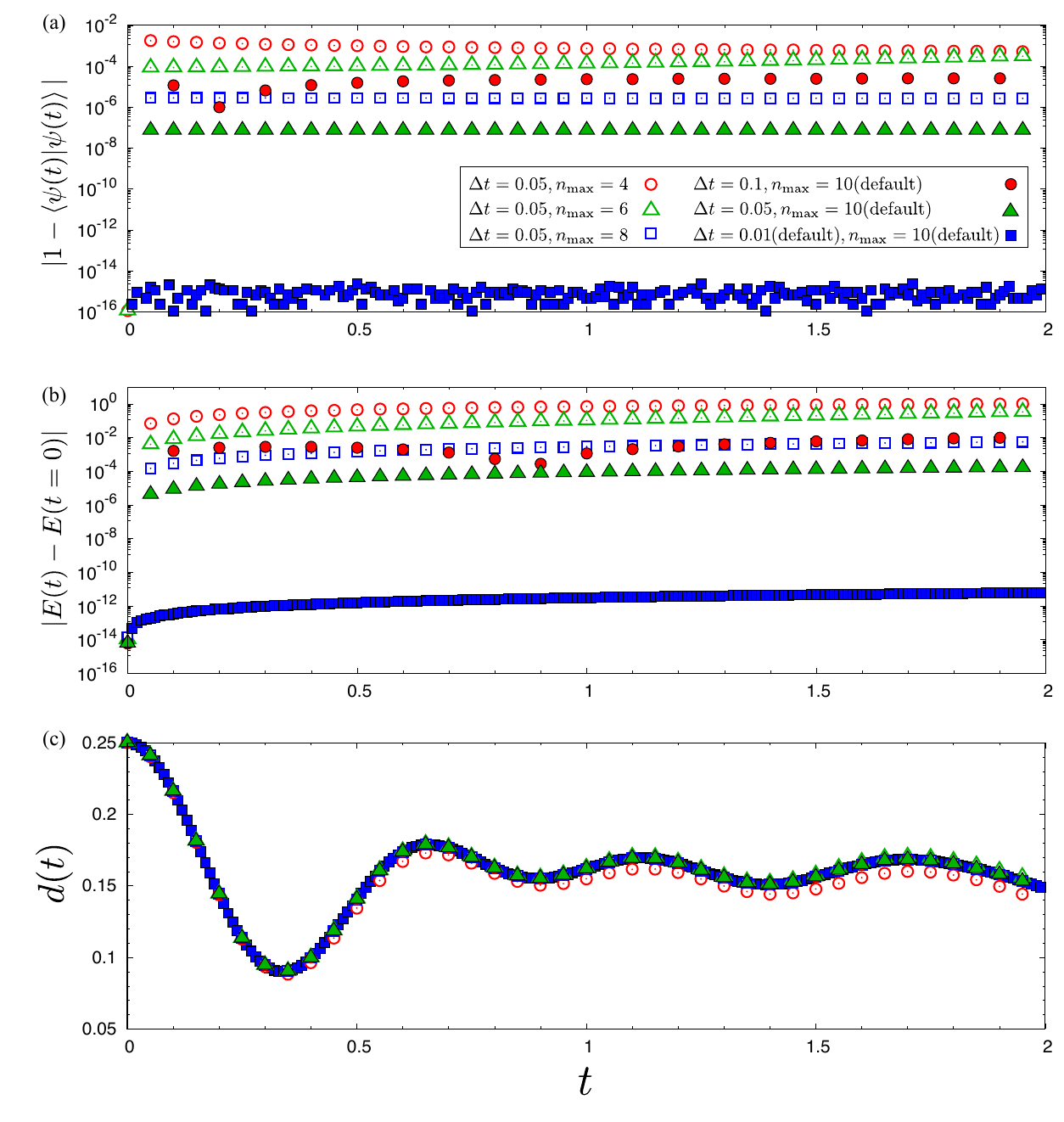}
\caption{
  Time-dependence of (a) norms, (b) energies, and (c) double occupancies during the sudden quench dynamics in the one-dimensional Hubbard model for $U_{0}/t_{\rm hop}=0$, $U_{\rm quench}/t_{\rm hop}=8$, and $N_{\rm s}=10$.
  Legends of panels (b) and (c) are the same as those of panel (a).
  The units of the energy and the time are $t_{\rm hop}$ and ${t^{-1}_{\rm hop}}$, respectively.
  In $\HPhi$, the defaults of $\Delta t$ and $n_{\rm max}$ are set to $0.01$ and $10$, respectively. 
  To enhance the visibility of the energy difference during the real-time evolution, the energy at $t=0$ outputted in the simulation with the default parameters is used as $E(t=0)$.
}
\label{te1}
\end{center}
\end{figure}

For the sudden quench simulation, the total energy $\ev{\mathcal{H}(t)}$ should be conserved after the interaction quench because the interaction term $U(t)$ defined in Eq. (\ref{Uquench}) does not have time-dependence for $t \geq 0$. 
In addition, the norm of the wave function $\braket{\Phi(t)}$ should not change during any unitary dynamics in quantum systems.
To check whether these properties are satisfied during $\HPhi$ simulation, the time-dependence of the norm and the total energy are outputted to \verb|Norm.dat| and \verb|SS.dat| in the \verb|output| directory, respectively.
Figure \ref{te1} shows the  $n_{\rm max}$ and $\Delta t$ dependence of these quantities and the double occupancy, 
\begin{equation}
d=\frac{1}{N_{\rm s}}\sum_i \ev*{n_{i\uparrow}n_{i\downarrow}}, 
\end{equation}
during a sudden quench.
We see that the conservation laws of the norm and the energy are satisfied for sufficiently large $n_\text{max}$ and small time step. The dynamics of double occupancy also converge under such conditions. The double occupancy shows a damped oscillation, which is often observed in sudden quench systems \cite{PhysRevLett.103.056403,PhysRevB.90.075117}.

\subsection{cTPQ}\label{sec:cTPQ}
\subsubsection{Basics of the TPQ method}
Here, we briefly summarize the basics of the TPQ states and
the difference between the microcanonical TPQ (mTPQ) state~\cite{Sugiura_PRL2012} 
and the canonical TPQ (cTPQ) state~\cite{Sugiura_PRL2013}. 
We note that the two methods are equivalent
if we take appropriate hyperparameters.
In $\HPhi$, both methods are implemented.
As we detail below,
because error estimation of the cTPQ method is straightforward 
compared with the mTPQ method, 
we recommend that users use the cTPQ method in actual calculations.

The TPQ state is defined in
\begin{align}
\ket{\Phi(\beta)}&=\exp[-\frac{\beta\hat{H}}{2}]\ket{\Phi_{\rm rand}}, \\
\ket{\Phi_{\rm rand}}&=\sum_{x}a_{x}\ket{x},
\end{align}
where $\beta$ represents the inverse temperature and
$\{\ket{x}\}$ is a basis set that spans the Hilbert space.
For example, for a Hamiltonian conserving the total particle number, $x$ may be the set of coordinates of particles, and for a quantum spin Hamiltonian conserving the total $S_z$, $x$ may be the set of 
spin positions in the real space.
In the actual calculations, we take the random number $a_{x}$ as uniformly distributed
on the $d_{\rm H}$-dimensional hypersphere, where $d_{\rm H}$ is the dimensionality of the 
Hilbert space of the given system.
If we can assume the initial random vectors uniformly 
include all the eigenstates $\ket{n}$, $\ket{\Phi_{\rm rand}}$
can be expressed as
\begin{align}
\ket{\Phi_{\rm rand}}=\sum_n a_{n}\ket{n},
\end{align}
where the weight $|a_{n}|$ can be approximated 
as $1/d_{\rm H}$.
Under this assumption, for example, the expectation value of a Hamiltonian is evaluated as 
\begin{align}
&\ev*{\hat{H}}=\frac{\ev{\hat{H}}{\Phi(\beta)}}{\braket{\Phi(\beta)}}
=\frac{\sum_{n}|a_{n}|^2E_{n}e^{-\beta E_{n}}}{\sum_{n}|a_{n}|^2e^{-\beta E_{n}}}\\
&\sim \frac{\sum_{n}E_{n}e^{-\beta E_{n}}}{\sum_{n}e^{-\beta E_{n}}}=E(\beta).
\end{align}
This is the ensemble average of the energy.
We note that this rough argument can be mathematically 
justified based on probability theory~\cite{Jin_JPSJ2021}.
As shown in the literature~\cite{Sugiura_PRL2013}, 
the expectation values with respect to the TPQ state can reproduce by the ensemble average even the physical quantities that do not commute with the Hamiltonian. 
Therefore, the main task of the TPQ calculation is to generate 
$\ket{\Phi(\beta)}$, in other words, to numerically apply
the exponential operator $e^{-\beta\hat{H}/2}$ to the random initial vector 
$\ket{\Phi_{\rm rand}}$.

According to Sugiura and Shimizu \cite{Sugiura_PRL2012,Sugiura_PRL2013}, there are two ways to calculate 
$\ket{\Phi(\beta)}$:
the microcanonical TPQ (mTPQ) state and the canonical TPQ (cTPQ) state.
In the following, we give their definitions. We rewrite the TPQ state by using the imaginary time-evolution operator $\hat{U}(\Delta \tau)$ as
\begin{align}
\exp\qty[-\frac{\beta}{2}\hat{H}]\ket{\Phi_{\rm rand}}\sim \hat{U}(\Delta \tau)^{k}\ket{\Phi_{\rm rand}},
\end{align}
where $\Delta\tau$ represents the slice width of the imaginary time evolution.
In the cTPQ state, the definition of $\hat{U}(\Delta \tau)$ is straightforward and given as
\begin{align}
&\exp[-\frac{\Delta\tau}{2}\hat{H}]\sim 
\sum_{n=0}^{n_{\rm max}}\frac{1}{n!}(-\frac{\Delta\tau}{2}\hat{H})^{n}\equiv \hat{U}_{\rm c}(\Delta \tau)\\
&\ket{\Phi_{\rm cTPQ}(\beta_{k})}=[\hat{U}_{\rm c}(\Delta \tau)]^{k}\ket{\Phi_{\rm rand}},\\
&\beta_{k}=k\Delta\tau.
\end{align}
In other words, we just expand $\hat{U}(\Delta\tau)$ with respect to $\Delta\tau$.
In the cTPQ calculation, $n_{\rm max}$ and $\Delta\tau$ are hyperparameters.
As we show later, $n_{\rm max}=10$ and $\Delta\tau=0.01J$ give well-converged
results in the one-dimensional antiferromagnetic Heisenberg model.
In $\HPhi$, we implement this Taylor expansion-based method
of constructing the cTPQ method because of its simplicity and generality. We note that the cTPQ state can also be constructed from the
mTPQ state as is shown in Ref.~\cite{Sugiura_PRL2013}.

In contrast, the definition of $\hat{U}(\Delta \tau)$ in the mTPQ state is
rather elaborate. In the mTPQ state, $\hat{U}(\Delta \tau)$ is defined as
\begin{align}
&\hat{U}_{\rm m}(\Delta \tau)\equiv(1-\frac{\Delta\tau}{2}\hat{H}),~\Delta\tau=\frac{2}{lN_{\rm s}},\\
&\ket{\Phi_{\rm mTPQ}(\beta_{k})}=[\hat{U}_{\rm m}(\Delta \tau)]^{k}\ket{\Phi_{\rm rand}},\\
&~\beta_{k}=\frac{k\Delta\tau}{1-u_{k}/l},\\
&u_{k}=\frac{1}{N_{\rm s}}\frac{\ev{\hat{U}_m(\Delta \tau)^{2k}\hat{H}}{\Phi_{\rm rand}}}{\ev{\hat{U}_m(\Delta \tau)^{2k}}{\Phi_{\rm rand}}},
\end{align}
where $l$ represents a constant. 
At first look, one might think that 
the mTPQ state is just a first-order cTPQ state ($n_{\rm max}=1$).
However, the definition of the inverse temperature is different.
In the denominator of $\beta_k$, $u_{k}/l$ is introduced in the mTPQ state.
This correction to $\beta$ may reduce the error caused by the
finite-order expansion of $\hat{U}(\Delta\tau)$.
From these definitions, it is obvious that the mTPQ states become equivalent to
the cTPQ state by taking the large $l$ limit, i.e., the small $\Delta\tau$ limit.

In the TPQ calculation, we prepare several different initial wave functions ($\ket*{\Phi_{\rm rand}^{n}}$,~$n=0,1,\dots,N-1$) and
evaluate the average values and the errors originating from the choice of the initial states.
In the procedure, we use the following bootstrap method:
\begin{enumerate}
\item 
Choosing $P$ samples from $N$ samples $M$ times with replacement.  We evaluate
the average values of the physical quantities $\hat{A}$ and the norm for each sample $m$ as follows:
\begin{align}
&\ket*{\Phi^{\mathcal{R}_{m}(p)}(\beta_{k})}=\hat{U}(\Delta\tau)^{k}\ket*{\Phi_{\rm rand}^{\mathcal{R}_{m}(p)}},\\
&A_{m}(\beta_{k})=\sum_{p=0}^{P-1}\ev*{\hat{A}}{\Phi^{\mathcal{R}_{m}(p)}(\beta_{k})},\\
&Z_{m}(\beta_{k})=\sum_{p=0}^{P-1}\braket*{\Phi^{\mathcal{R}_{m}(p)}(\beta_{k})},\\
&\ev*{A_{m}(\beta_{k})}=\frac{A_{m}(\beta_{k})}{Z_{m}(\beta_{k})},
\end{align}
where $\mathcal{R}_{m}(p)$ is a randomly chosen number in $P$ samples and
$m$ represents the index of the bootstrap sampling ($m=0,1,\dots,M-1$). 
We note that it is acceptable to duplicate each sample, i.e.,
$\mathcal{R}_{m}(p)=\mathcal{R}_{m}(q)$ for $p\neq q$ is acceptable.
\item Then, we evaluate the average values and the error of the physical quantities as
\begin{align}
&\bar{A}(\beta_{k})=\frac{1}{M}\sum_{m=0}^{M-1}\ev*{A_{m}(\beta_{k})},\\
&\sigma\qty[{A}(\beta_{k})]^2=\frac{1}{M-1}\sum_{m=0}^{M-1}\qty[A_{m}(\beta_{k})^2-\bar{A}(\beta_{k}^2)].
\end{align}
\end{enumerate}
In the cTPQ method, $\beta_{k}$ is the same for each initial state. Thus,
the bootstrap sampling can be done by the above procedure.
In contrast,
in the mTPQ method, $\beta_{k}$ becomes different for each initial state although
its difference is small in many cases. 
Therefore, we should do interpolation or extrapolation of the physical quantities
to do the bootstrap sampling for the mTPQ states.
However, this may make the error estimation for the mTPQ method complicated.
Therefore, we recommend that users use the cTPQ method to be able to estimate the errors accurately. 
Nevertheless, the mTPQ method is still
a useful and simple method for capturing general trends of target
systems because the mTPQ requires only one matrix-vector operation.
 
Here, we explain the technical details of the TPQ calculations.
Since $Z_{m}(\beta_{k})$ becomes exponentially large by lowering the temperature,
it is necessary to normalize $Z_{m}(\beta_{k})$ with a proper factor.
A simple way to conduct normalization is to use
$\mathcal{W}(\beta_{k})=\braket*{\Phi^{0}(\beta_{k})}$, i.e., 
the squared norm at the zeroth initial state.
Under normalization, the physical quantities are given as
\begin{align}
&\tilde{Z}_{m}(\beta_{k})=\frac{Z_{m}(\beta_{k})}{\mathcal{W}(\beta_{k})},~\tilde{A}_{m}(\beta_{k})=\frac{A_{m}(\beta_{k})}{\mathcal{W}(\beta_{k})},\\
&\ev*{A_{m}(\beta_{k})}=\frac{\tilde{A}_{m}(\beta_{k})}{\tilde{Z}_{m}(\beta_{k})}.
\end{align}
In the actual calculations, we obtain $\tilde{A}_{m}(\beta_{k})$ and
$\tilde{Z}_{m}(\beta_{k})$, then perform the bootstrap calculations.
We note that the information of the norm is 
output in \verb|Norm_n.dat| (n represents the number of the initial states) in $\HPhi$.
The squared norm of $\ket{\Psi(\beta_k)}$ is defined as
\begin{align}
N_{k} = \braket*{\Psi(\beta_k)}.
\end{align}
In \verb|Norm_m.dat|, the relative squared norm $D_{k}$ is outputted
since the value of $N_{k}$ also becomes exponentially large.
The definition of the relative norm $D_{k}$ is 
\begin{align}
D_{0}&= N_{0} = \braket*{\Phi_{\rm rand}},\\
D_{k}&=\frac{N_{k}}{N_{k-1}}~(\text{for $k>0$}).
\end{align}
From the relative squared norm, we can calculate the norm as
\begin{align}
N_{k}&=\prod_{l=0}^{k}D_{k}.
\end{align}

\begin{figure}[t] 
\begin{center} 
\includegraphics[width=0.48 \textwidth]{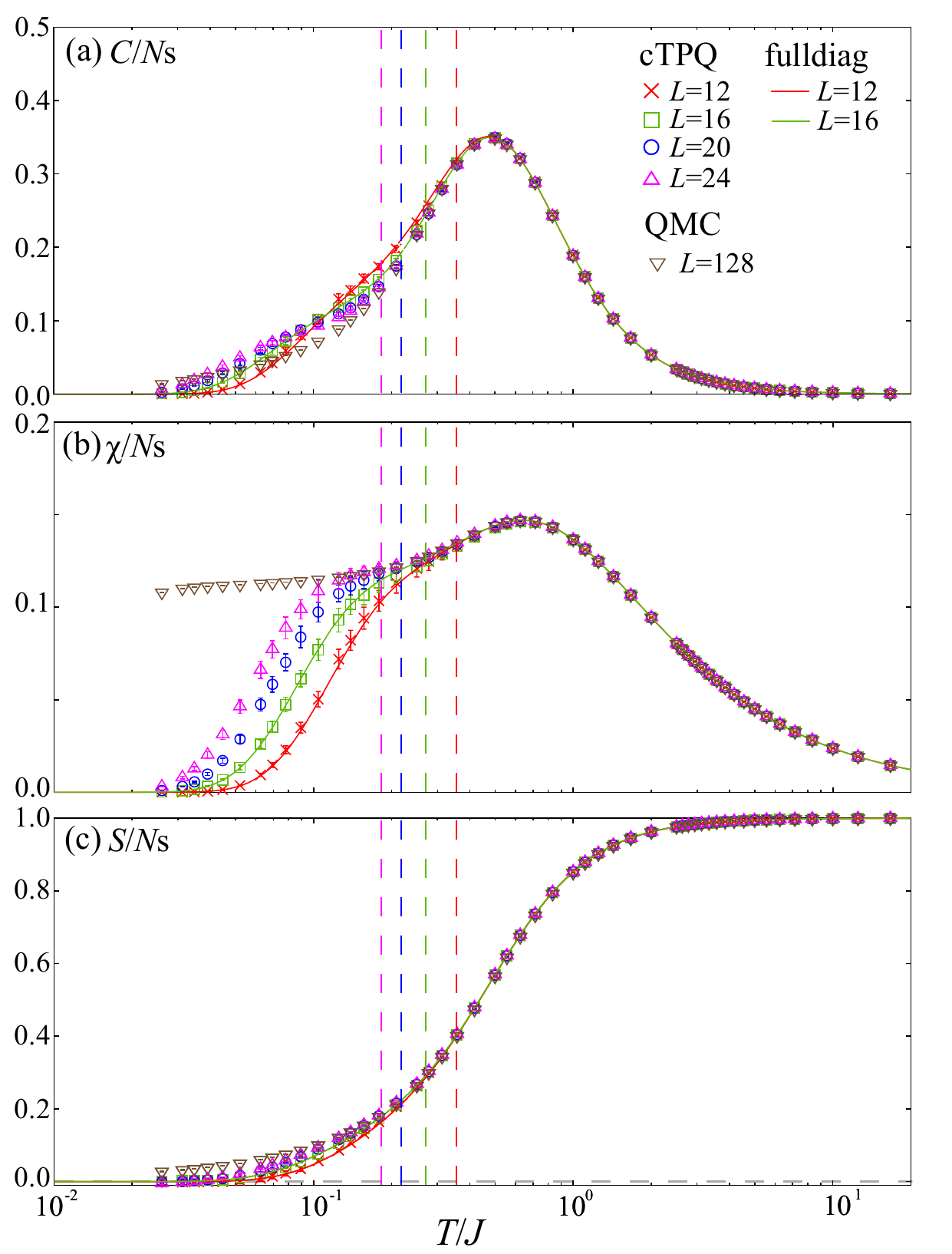}
\caption{Temperature dependence of 
(a) specific heat $C$, (b) the uniform magnetic susceptibility $\chi$ , and (c) the entropy $S$. The results of the cTPQ (full diagonalization)
calculations are shown by symbols (lines).
We note $\chi$ is calculated from the fluctuation of total $S_{z}$ ($S$) 
in the cTPQ method (full diagonalization).
We also show the results by the quantum Monte Carlo (QMC) method
for large system size ($L=128$). The QMC calculation is done using DSQSS v2.0.4~\cite{dsqss,Motoyama_CPC2021}.
Error bars arising from Monte Carlo samplings are shown, but all of them are smaller than the symbol sizes.
The vertical broken lines show the singlet-triplet gap for $L=12,16$, and $20$.
}
\label{cTPQ}
\end{center}
\end{figure}

\subsubsection{Benchmark for the Heisenberg model}
Here, we show a benchmarking result of the cTPQ method for the $L$-site one-dimensional
Heisenberg model ($L=12,16,20,24$). 
A sample script is available in Ref.~\cite{repo}.
We calculate several physical quantities, such as 
the specific heat $C$, the entropy $S$, and the
magnetic susceptibility $\chi$, which are defined as
\begin{align}
C&=\frac{\ev*{\hat{H}^2}-\ev*{\hat{H}}^2}{T^{2}}, \\
S&=\frac{\log{Z}+\frac{\ev*{\hat{H}}}{T}}{\log{d_{\rm H}}},\\
\chi_{z}&=\frac{\ev*{\hat{S}_{z}^2}-\ev*{\hat{S}_{z}}^2}{T}
=\frac{\ev*{\hat{\vec{S}}^2}-\ev*{\hat{\vec{S}}}^2}{3T},\\
\hat{S}_{z}&=\sum_{i}\hat{S}_{iz},\\
\hat{\vec{S}}&=\sum_{i}(\hat{S}_{ix},\hat{S}_{iy},\hat{S}_{iz}),
\end{align}
where $d_{\rm H}$ is the Hilbert dimensions of the given systems
($d_{\rm H}=2^{L}$ for the one-dimensional Heisenberg model).
In the cTPQ calculation, 
we take $n_{\rm max}=6$ and $\Delta \tau=0.02$.
In the bootstrap sampling,
we take $N$ independent initial states, 
and choose $P$ samples $M$ times from $N$ samples with replacement. 
In this calculation, we take $P=N=200$, and $M=N/2=100$.

As shown in Fig.~\ref{cTPQ},
the cTPQ method well reproduces the results obtained by the 
full diagonalization for $L=12$ and $L=16$.
Although the total dimension is not so large for $L=12$ ($d_{\rm H}=2^{12}=4096$),
the results of the cTPQ methods are consistent with those of the full diagonalization
within the error. 
We note that the difference between physical quantities calculated by the cTPQ method and
those by the canonical ensemble becomes exponentially small as a function of the Hilbert dimension $d_{H}$~\cite{Sugiura_PRL2013}.
We also compare the results by the quantum Monte Carlo (QMC) method for large system size ($L=128$) with those by the cTPQ method.
Above the singlet-triplet gap arising from finite-size effects (shown by the broken lines),
we find that the cTPQ method well reproduces the result by QMC, which
can be regarded as the thermodynamic-limit values in the calculated temperature region. These results demonstrate that the cTPQ method is a reliable method for 
calculating the finite-temperature properties of 
quantum-many body systems.

\section{Update of Standard mode}
\label{sec:standard}
\subsection{Important keywords}
To simplify the input format in $\HPhi$, we prepare Standard mode,
which enables us to perform calculations for 
standard quantum lattice models.
We first give an overview of the basic keywords used in Standard mode; 
\verb|model|, \verb|method|, and \verb|lattice|.
Here, we denote the keywords used in Standard mode
by the typewriter font such as {\tt model}.
We list the available models, methods, and lattices in Table~\ref{table:Keywords}.

\begin{table*}[t!]
\caption{Main keywords used in Standard mode of $\HPhi$. 
}
\begin{tabular}{ll}
\hline 
Model name & Explanation \\
\hline 
Spin       & Spin-$S$ Heisenberg model with total $S_{z}$ conservation\\ 
SpinGC     & Spin-$S$ Heisenberg model {\it without} total $S_{z}$ conservation\\
SpinCMA    & Continuous memory access method for SpinGC systems~\cite{CMA} \\
Hubbard    & Hubbard model with particle $N$ conservation and with/without total $S_{z}$ conservation\\ 
HubbardGC  & Hubbard model {\it without} particle $N$ conservation and total $S_{z}$ conservation\\  
Kondo      & Kondo lattice model with particle $N$ conservation and total $S_{z}$ conservation\\
KondoGC    & Kondo lattice model {\it without} total $S_{z}$ and particle $N$ conservation  \\
\hline
\\
\hline 
Method name & Explanation \\
\hline
FullDiag         & Full diagonalization using LAPACK, ScaLAPACK, and MAGMA           \\  
Lanczos          & Ground-state and spectrum calculations using the Lanczos method \\ 
CG               & Ground-state (spectrum) calculations using the LOBPCG (shifted-Krylov) method\\ 
TPQ              & mTPQ method for finite-temperature calculation\\
cTPQ             & cTPQ method for finite-temperature calculation \\ 
Time-Evolution   & Real-time evolution\\
\hline
\\
\hline
Lattice name & Explanation \\
\hline
Chain         & One-dimensional chain lattice \\
Ladder        & $N$-leg ladder \\ 
Square        & Two-dimensional square lattice\\
Triangular    & Two-dimensional triangular lattice\\
Honeycomb     & Two-dimensional honeycomb structure\\
Kagome        & Two-dimensional kagome lattice\\
Cubic         & Three-dimensional monoatomic simple cubic lattice\\
FCO           & Three-dimensional monoatomic face-centered orthorhombic lattice\\
Pyrochlore    & Three-dimensional pyrochlore structure\\
Wannier90     & Input from the transfer integrals and interactions in the Wannier90 format\\
\hline
\end{tabular}
\label{table:Keywords}
\end{table*}

\subsubsection{Remarks on the keyword model}
Below, we give several remarks on the models.
Regarding spin models, we can treat general spin-$S$ Heisenberg models other than
the $S=1/2$ one. For example, by specifying
\begin{verbatim}
2S=2
\end{verbatim}
in Standard mode, one can treat the $S=1$ Heisenberg model.
In Expert mode, by preparing the amplitude of spin at each site
in \verb|localspin.def|,
it is possible to treat mixed spin systems, e.g., spin systems
with $S=1/2$ and $S=1$ spins.

If one does not specify the total $S_{z}$ in \verb|model="Hubbard"|,
one can treat the Hubbard-type model without total $S_{z}$ conservation.
Although \verb|HubbardGC| allows us to treat the systems without particle-number conservation,
the one-body and two-body terms that break the particle-number conservation 
 (anomalous terms) are not implemented in either Standard mode or Expert mode of $\HPhi$.

In Standard mode, we can treat the Kondo-lattice model, which consists
of $S=1/2$ spins and itinerant electrons systems.
In Expert mode, we can treat the general-$S$ Kondo-lattice model
by changing \verb|localspin.def| properly.

\subsubsection{Remarks on the keyword method}
We first note that \verb|Lanczos| and \verb|CG| are used
both for ground-state calculations and spectrum calculations.
In the spectrum calculations,  \verb|Lanczos| (\verb|CG|)
refers to the spectrum calculations by 
the continued-fraction expansion (shifted-Krylov method).
\verb|TPQ| means the micro-canonical TPQ method. 
However, as we mentioned above,
the \verb|cTPQ| method is recommended for the actual calculations
since the procedure of the error estimation is straightforward.

\subsubsection{Remarks on the keyword lattice}
In $\HPhi$, several typical lattice structures are prepared for use in Standard mode. 
Keywords \verb|L|, \verb|W|, and \verb|Height| represent the lengths of the system sites in the $x$, $y$, and $z$ directions, respectively. 
Unnecessary lengths in higher dimensions 
(e.g., \verb|Height| in a two-dimensional lattice) can be omitted.
Note that \verb|H| is a keyword for a magnetic field.
Standard mode can also treat more complicated lattice structures such as
a tilted lattice. 
Here, we give the specifications of the lattice structure in Standard mode:
Let $\vec{a}_{0}$, $\vec{a}_{1}$, and $\vec{a}_{2}$ be unit vectors in real space.
The position of the unit cell is described by
\begin{align}
\vec{R}=r_{0}\vec{a}_{0}+r_{1}\vec{a}_{1}+r_{2}\vec{a}_{2},
\end{align}
where $r_{i}$ ($i=0,1,2$) is an integer.
We consider the finite-size lattice (supercell) spanned by the
linear combination of these unit vectors.
The linear combination is defined as
\begin{align}
&\begin{pmatrix}
\vec{A}_{0} \\
\vec{A}_{1} \\
\vec{A}_{2} 
\end{pmatrix}
=N
\begin{pmatrix}
\vec{a}_{0} \\
\vec{a}_{1} \\
\vec{a}_{2} 
\end{pmatrix},\\
&N=
\begin{pmatrix}
N_{00}~& N_{01}~& N_{02} \\
N_{10}~& N_{11}~& N_{12} \\
N_{20}~& N_{21}~& N_{22} 
\end{pmatrix},
\end{align}
where $N_{ij}$ ($i,j=0,1,2$) is an integer.
$N_{i0}$, $N_{i1}$, $N_{i2}$ 
are specified by {\tt a$i$W}, {\tt a$i$L}, and {\tt a$i$H} ($i=0,1,2$).
\begin{figure}[t] 
    \begin{center} 
    \includegraphics[width=0.45 \textwidth]{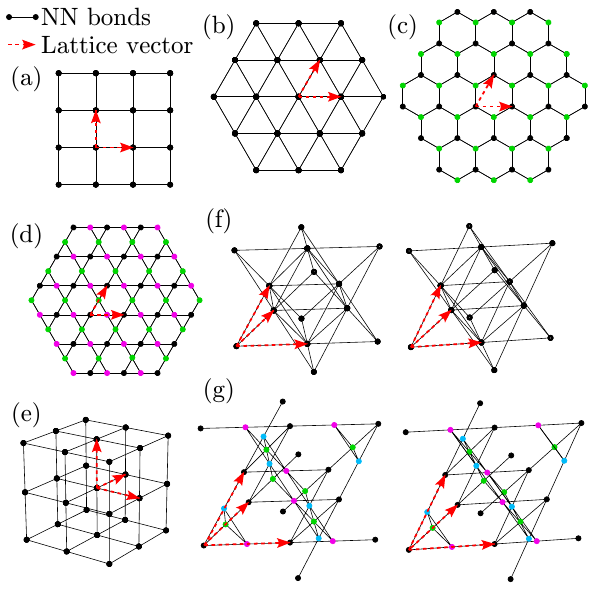}
    \caption{
        Two- and three-dimensional predefined lattices: (a) square lattice, (b) triangular lattice, (c) honeycomb lattice, (d) kagome lattice (e) simple cubic lattice, (f) face-centered cubic lattice, (g) pyrochlore structure.
        (f) and (g) are cross-eyed stereograms.
    }
    \label{fig:lattice}
   \end{center}
\end{figure}
Unit vectors and the Wyckoff positions for the two- and three-dimensional lattices defined in Standard mode are summarized below (see also Fig. \ref{fig:lattice}):
\begin{itemize}
    \item Square lattice [Fig. \ref{fig:lattice} (a)]
    
    The unit lattice vectors are
    \begin{align}
        \vec{a}_0 &= (1, 0, 0), \\
        \vec{a}_1 &= (0, 1, 0).
    \end{align}
    \item Triangular lattice, honeycomb lattice, kagome lattice [Figs. \ref{fig:lattice} (b, c, d)]

    The unit lattice vectors are
    \begin{align}
        \vec{a}_0 &= (1, 0, 0), \\
        \vec{a}_1 &= \left(\frac{1}{2}, \frac{\sqrt{3}}{2}, 0\right).
    \end{align}
    The honeycomb lattice has two sites in a unit cell,
    \begin{align}
        \vec{0}, \quad \frac{\vec{a}_0}{3} + \frac{\vec{a}_1}{3},
    \end{align}
    while the kagome lattice has the three sites
    \begin{align}
        \vec{0}, \quad \frac{\vec{a}_0}{2}, \quad \frac{\vec{a}_1}{2}.
    \end{align}

    \item Simple cubic lattice [Fig. \ref{fig:lattice} (e)]

    The unit lattice vectors are
    \begin{align}
        \vec{a}_0 &= (1, 0, 0), \\
        \vec{a}_1 &= (0, 1, 0), \\
        \vec{a}_2 &= (0, 0, 1).
    \end{align}
    
    \item Face-centered cubic lattice, pyrochlore structure [Figs. \ref{fig:lattice} (f, g)]

    The unit lattice vectors are
    \begin{align}
        \vec{a}_0 &= \left(0, \frac{1}{2}, \frac{1}{2}\right), \\
        \vec{a}_1 &= \left(\frac{1}{2}, 0, \frac{1}{2}\right), \\
        \vec{a}_2 &= \left(\frac{1}{2}, \frac{1}{2}, 0\right).
    \end{align}
    The pyrochlore structure has the following four sites in a unit cell:
    \begin{align}
        \vec{0}, \quad \frac{\vec{a}_0}{2}, \quad \frac{\vec{a}_1}{2}, \quad \frac{\vec{a}_2}{2}.
    \end{align}
   
\end{itemize}

We also note that the wave numbers, which are compatible with
the above real-space lattice, can be obtained as follows.
The reciprocal vectors are defined as
\begin{align}
\vec{b}_{0}&=2\pi\frac{\vec{a}_{1}\times\vec{a}_{2}}{\vec{a}_{0}\cdot(\vec{a}_{1}\times\vec{a}_{2})}, \\
\vec{b}_{1}&=2\pi\frac{\vec{a}_{2}\times\vec{a}_{0}}{\vec{a}_{1}\cdot(\vec{a}_{2}\times\vec{a}_{0})},\\
\vec{b}_{2}&=2\pi\frac{\vec{a}_{0}\times\vec{a}_{1}}{\vec{a}_{2}\cdot(\vec{a}_{0}\times\vec{a}_{1})}.
\end{align}
The position of the unit cell in the reciprocal space is defined as
\begin{align}
\vec{K}=k_{0}\vec{b}_{0}+k_{1}\vec{b}_{1}+k_{2}\vec{b}_{2}.
\end{align}
Since $\vec{R}$ and $\vec{R}+\vec{A}_{i}$ are equivalent,
the following relation holds for the commensurate wave vector:
\begin{align}
e^{i\vec{K}_{\rm com}\cdot\vec{R}}=e^{i\vec{K}_{\rm com}\cdot(\vec{R}+\vec{A}_{i})}=1.
\end{align}
In general, the commensurate wave vectors are given by
\begin{align}
&\vec{K}_{\rm com}=m_{0}\vec{B}_{0}+m_{1}\vec{B}_{1}+m_{2}\vec{B}_{2},\\
&\begin{pmatrix}
\vec{B}_{0}, 
\vec{B}_{1},
\vec{B}_{2} 
\end{pmatrix}
=
\begin{pmatrix}
\vec{b}_{0},
\vec{b}_{1}, 
\vec{b}_{2} 
\end{pmatrix}
N^{-1},
\end{align}
where $m_{i}$ is an integer and $N$ is the total number of unit cells.

\section{Support for Wannier90 format}
\label{sec:wan90}

$\HPhi$  supports reading input files of transfer integrals, direct-Coulomb and exchange integrals written in Wannier90 format~\cite{wan90_HP,Pizzi_2020JPC} by setting \verb|mode="wannier90"|. 
Using this mode, the following Hamiltonian can be defined:
\begin{align}
&\mathcal{H}=\mathcal{H}_{\rm 1body}+\mathcal{H}_{\rm 2body} \label{eq:ham_general}\\
&\mathcal{H}_{\rm 1body}=\sum_{m,n, i, j,\sigma} t_{mn}({\bf R}_{ij}) c_{im \sigma}^{\dagger} c_{jn \sigma} \label{eq:ham_onebody} \\
&\mathcal{H}_{\rm 2body}=\mathcal{H}_{U}+\mathcal{H}_{J}\\
&\mathcal{H}_{U}=\sum_{i,m} U_{mm}({\bf 0})n_{im,\uparrow}n_{im, \downarrow}\nonumber\\
&+\sum_{(i,m)<(j,n)}U_{mn}({\bf R}_{ij})n_{im}n_{jn}\nonumber\\
&\mathcal{H}_{J}=-\sum_{(i,m)<(j,n)}J_{mn}({\bf R}_{ij})(n_{im, \uparrow}n_{jn,\uparrow}+n_{im, \downarrow}n_{jn,\downarrow}) \nonumber\\
& + \sum_{(i,m)<(j,n)}J_{mn}({\bf R}_{ij})(c_{im, \uparrow}^{\dagger}c_{jn,\uparrow}c_{jn,\downarrow}^{\dagger}c_{im,\downarrow}+{\rm h.c.}) \nonumber\\
& + \sum_{(i,m)<(j,n)}J_{mn}({\bf R}_{ij}) (c_{im, \uparrow}^{\dagger}c_{jn,\uparrow}c_{im,\downarrow}^{\dagger}c_{jn,\downarrow} + {\rm h.c.} ),
\end{align}
where \({\bf R}_{ij}={\bf R}_j-{\bf R}_i\). Here, $t_{mn}({\bf R}_{ij})$, $U_{mn}({\bf R}_{ij})$, $J_{mn}({\bf R}_{ij})$ are a transfer integral, Coulomb and exchange integrals 
between $m$-orbit at ${\bf R}_i$ and $n$-orbit at ${\bf R}_j$, 
and can be specified by the input files named as \verb|zvo_hr.dat|, \verb|zvo_ur.dat|, and \verb|zvo_jr.dat| in the \verb|dir-model| directory, respectively. 
These terms correspond to 
the on-site Coulomb interaction (\verb|CoulombIntra|), 
the off-site Coulomb interaction (\verb|CoulombInter|),
the Hund couplings (\verb|Hund|),
the exchange couplings (\verb|Exchange|), and
the pairhopping terms (\verb|Pairhopping|).
Here, the geometry of the lattice can be defined by the input file 
named as \verb|zvo_geom.dat|.  Below, we detail the format of these files.

Information regarding the lattice structure is specified in \verb|zvo_geom.dat| and is defined as follows:
\begin{verbatim}
   a0x  a0y  a0z
   a1x  a1y  a1z
   a2x  a2y  a2z
   Nwan
   C0x  C0y  C0z
   C1x  C1y  C1z
   ...
\end{verbatim}
In the first three lines, the components of the unit vectors ($\vec{a}_{i}$) are specified. 
The number of Wannier orbitals ($N_{\rm wan}$) is defined in the fourth line.
After the fifth line, the center positions of the Wannier orbitals ($\vec{C}_{i}$) are given in fractional coordinates.

The parameters of the Hamiltonians such as the transfer integrals (\verb|zvo_hr.dat|),
the Coulomb interactions (\verb|zvo_ur.dat|), and the exchange interactions (\verb|zvo_jr.dat|)
can be specified in the Wannier90 format.
We note that $\HPhi$ only supports the {\it reducible} representation, i.e., 
that each translation vector is regarded as independent.
Below, we summarize the structure of the Wannier90 format with a reducible representation for transfer integrals.
\begin{verbatim}
  Comment
  Nwan
  Nscel
  1 ... 1
  1 ... 1
  Rx  Ry  Rz  m  n  Re[tmn(R)] Im[tmn(R)]
\end{verbatim}
In the Wannier90 format, the first line can be any text.
The number of Wannier orbitals ($N_{\rm wan}$) is given in the second line,
the number of supercells ($N_{\rm scel}$) is specified by the third line,
the degeneracy of each Wigner-Seitz grid point is given in the fourth line~\footnote{Since we do not use the information of the degeneracy in $\HPhi$, we always set the degeneracy as 1 for simplicity.}, 
and the information on the parameters of Hamiltonians is given in the following line.
We note that the number of degeneracies should be arranged as 15 pieces in each line. 
The position vector is represented by 
$\vec{R}=(R_x,R_y,R_z)$ and orbital indices are shown by $m$ and $n$.
Each line corresponds to the following transfer integrals $t_{mn}(\vec{R}_{ij}=\vec{R}-\vec{0}$) 
[$\vec{R}_{i}=\vec{0}$ and $\vec{R}_{j}=\vec{R}$], 
defined in Eq.~(\ref{eq:ham_onebody}).
Using the same format, direct Coulomb integrals $U_{mn}(\vec{R})$ and direct exchange integrals $J_{mn}(\vec{R})$ can be defined by \verb|zvo_ur.dat| and \verb|zvo_jr.dat|, respectively.

\begin{figure}[t] 
\begin{center} 
\includegraphics[width=0.45 \textwidth]{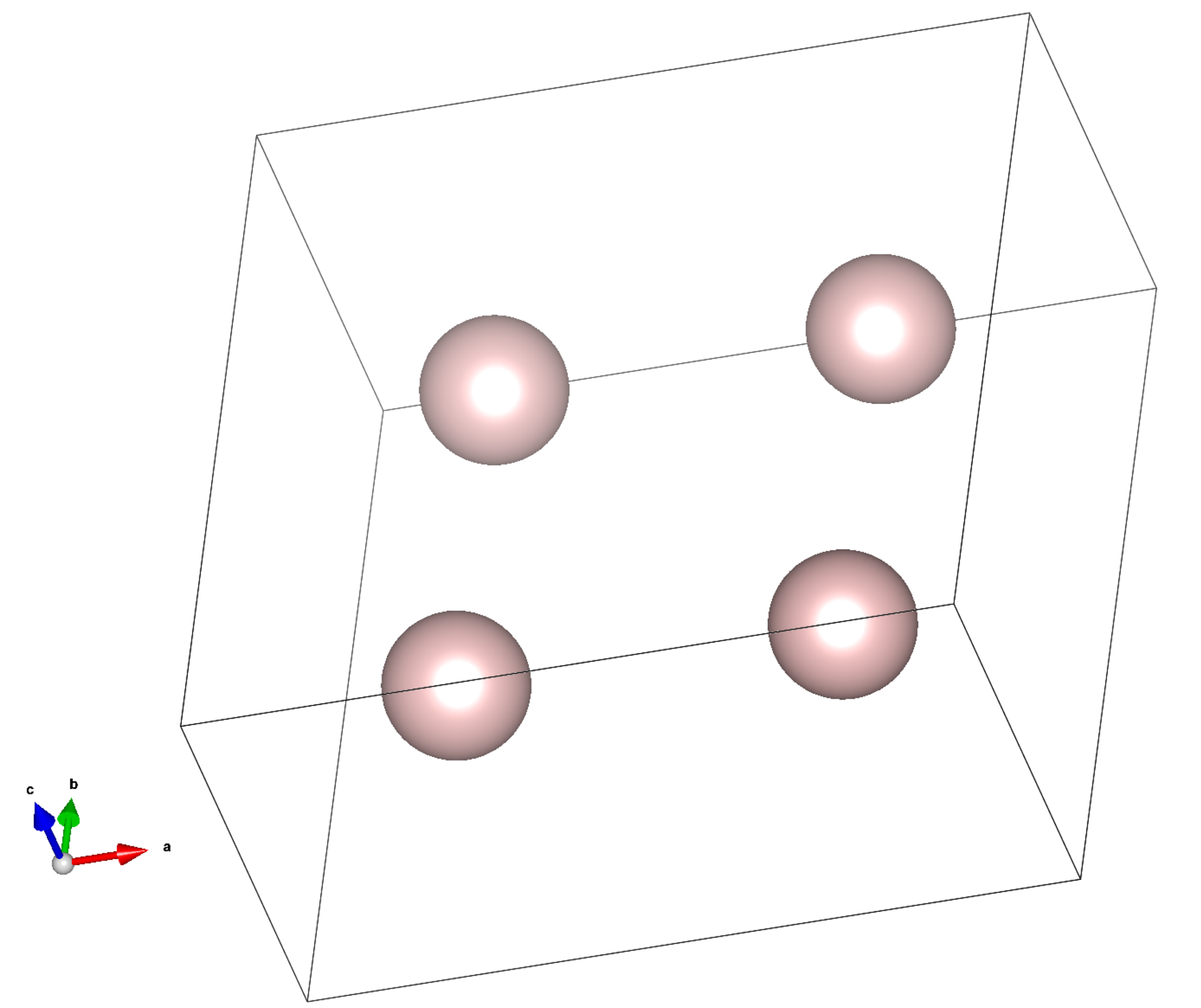}
\caption{
Visualization of the xsf file for a $2\times2$ square lattice by VESTA~\cite{VESTA}.
}
\label{lattice2by2}
\end{center}
\end{figure}

As a simple example, we show the input files for the two-dimensional Hubbard model with $t=1$, $U=4$ on a square lattice. The lattice geometry is given by  \verb|zvo_geom.dat|.
\begin{verbatim}
  1.0000000000   0.0000000000   0.0000000000
  0.0000000000   1.0000000000   0.0000000000
  0.0000000000   0.0000000000   1.0000000000
  1
  0.5000000000 0.5000000000 0.5000000000
\end{verbatim}
In this case, unit vectors are given by
\begin{align*}
&\vec{a}_{0}=(1,0,0),\\
&\vec{a}_{1}=(0,1,0),\\
&\vec{a}_{2}=(0,0,1).
\end{align*}
In the fourth line, the number of Wannier orbitals ($N_{\rm wan}$) is outputted (in this case, $N_{\rm wan}=1$). The center position of the Wannier orbital is defined as
\begin{align}
&\vec{C}_{\rm 0}=0.5\vec{a}_{0}+0.5\vec{a}_{1}+0.5\vec{a}_{2}.
\end{align}
Next, the nearest-neighbor transfer integrals are specified by \verb|zvo_hr.dat| given as follows:
\begin{verbatim}
wannier90 format
         1
         9
    1    1    1    1    1    1    1    1    1
   -1   -1    0    1    1    0.0  0.0
   -1    0    0    1    1   -1.0  0.0
   -1    1    0    1    1    0.0  0.0
    0   -1    0    1    1   -1.0  0.0
    0    0    0    1    1    0.0  0.0
    0    1    0    1    1   -1.0  0.0
    1   -1    0    1    1    0.0  0.0
    1    0    0    1    1   -1.0  0.0
    1    1    0    1    1    0.0  0.0
\end{verbatim}
Note that the transfer integral is defined by $-t$ in $\HPhi$. 
Thus, -1 in \verb|zvo_hr.dat| means $t=1$.
Finally, the on-site Coulomb interactions are specified by \verb|zvo_ur.dat| given as follows: 
\begin{verbatim}
wannier90 format
         1
         9
    1    1    1    1    1    1    1    1    1
   -1   -1    0    1    1    0.0  0.0
   -1    0    0    1    1    0.0  0.0
   -1    1    0    1    1    0.0  0.0
    0   -1    0    1    1    0.0  0.0
    0    0    0    1    1    4.0  0.0
    0    1    0    1    1    0.0  0.0
    1   -1    0    1    1    0.0  0.0
    1    0    0    1    1    0.0  0.0
    1    1    0    1    1    0.0  0.0
\end{verbatim}
The lattice size is specified using the input file of the standard mode of $\HPhi$.
Below, an example of the input file of the standard mode of $\HPhi$ is shown:
\begin{verbatim}
model = "Hubbard"
lattice = "Wannier90"
W = 2
L = 2
method = "CG"
2Sz = 0
nelec = 4
exct = 1
\end{verbatim}
In this case, the $N_s = 2\times2$ lattice is defined.
The lattice structure is output in the xsf format (\verb|lattice.xsf|).
One can confirm the lattice structure using visualization software such as VESTA~\cite{VESTA}. 
In Fig.~\ref{lattice2by2}, 
we show the lattice structure visualized by VESTA.

\section{Interface to RESPACK}
\label{sec:respack}
$\HPhi$  can make input files from the $ab$ $initio$ low-energy
effective Hamiltonians derived by RESPACK, which are specified
in the Wannier90 format. In this section,
we give an overview of the derivation
of the $ab$ $initio$ effective Hamiltonians
and the format of output/input files.

\begin{table*}[htb!]
\caption{Main output files by RESPACK. We assume the prefix of the file name is zvo.
Note that $\HPhi$ and mVMC support the Wannier90 format with the reducible representation.}
\begin{tabular}{ll}
\hline 
File name & Explanation \\
\hline 
zvo\_geom.dat & Information of the lattice unit vectors and positions of the Wannier centers \\
zvo\_hr.dat   & Transfer integrals    $t_{mn}(\vec{R}_{ij})$  (Wannier90 format) \\
zvo\_ur.dat   & Coulomb  interactions $U_{mn}(\vec{R}_{ij})$  (Wannier90 format)\\ 
zvo\_jr.dat   & Exchange interactions $J_{mn}(\vec{R}_{ij})$  (Wannier90 format)\\ 
zvo\_dr.dat   & Expectation values by Kohn-Sham Hamiltonians $D_{mn}(\vec{R}_{ij})$ (Wannier90 format)\\
\hline
\end{tabular}
\label{table:RESPACK}
\end{table*}

\subsection{Derivation of $ab$ $initio$ low-energy effective Hamiltonians}
We explain how to derive the effective Hamiltonians
from $ab$ $initio$ calculations.
We first obtain the global band structures for the target compounds 
using the software packages for $ab$ $initio$ calculations
such as Quantum ESPRESSO~\cite{QE,QE_HP} and xTAPP~\cite{xTAPP}.
Then, we construct Wannier functions that
describe the low-energy degrees of freedom of the target compounds.
Finally, using the Wannier functions,
we obtain the low-energy effective Hamiltonians defined in Eq.~(\ref{eq:ham_general}).

In RESPACK, the transfer integral $t_{mn}(\vec{R}_{ij})$ is  evaluated as
\begin{align}
t_{mn}(\vec{R}_{ij})=
\int_V d{\bf r}d{\bf r'}
w_{m{\bf 0}}^*({\bf r}) 
H_{0}({\bf r,r'})
w_{n{\bf R}}({\bf r'}).
\end{align}
Here, \(V\) is the volume of the crystal, \(w_ {i {\bf R}}({\bf r})\) is the $i$th Wannier 
function at the \(\bf R\)th cell, and $H_{0}(\vec{r}, \vec{r}^{\prime})$ is the Kohn-Sham Hamiltonian. 
The static screened direct integrals \(U_{mn}({\bf R})\) 
and the static screened exchanged integrals \(J_{mn}({\bf R})\) are given as follows:
\begin{align*}
&U_{mn}({\bf R})\\
&=\int_V d{\bf r}d{\bf r'}
w_{m{\bf 0}}^*({\bf r}) w_{m{\bf 0}}({\bf r})
W({\bf r,r'})
w_{n{\bf R}}^*({\bf r'}) w_{n{\bf R}}({\bf r'}),\nonumber\\
&J_{mn}({\bf R})\\
&=\int_V d{\bf r}d{\bf r'}
w_{m{\bf 0}}^*({\bf r}) w_{n{\bf R}}({\bf r})
W({\bf r,r'})
w_{n{\bf R}}^*({\bf r'}) w_{m{\bf 0}}({\bf r'}).
\end{align*}
Here, \(W({\bf r,r'})\) is the screened Coulomb interactions.
The information on lattice geometry, the positions of the Wannier centers, and the expectation values 
of the Kohn-Sham Hamiltonian are also outputted.
We summarize the output files by RESPACK in Table~\ref{table:RESPACK}.
The parameters of the Hamiltonian and one-body expectation values are output in \verb|zvo_hr.dat|,  \verb|zvo_ur.dat|, \verb|zvo_jr.dat| and \verb|zvo_dr.dat|, respectively.

We note that the following elimination of the double counting
of the chemical potential is necessary for multi-orbital systems~\cite{Misawa_JPSJ2011,Seo_JPSJ2013}:
\begin{align}
\mathcal{H} &=
\sum_{m, i, \sigma}
\left[t_{mm}(0) - t_{mm}^{\rm DC}(0)\right] c_{im \sigma}^{\dagger} c_{im \sigma},
\end{align}
where  \(t_{mm}^{\rm DC}(0)\) is the one-body correction term given as:
\begin{align}
&t_{mm}^{\rm DC}({\bf 0})= \alpha U_{mm}({\bf 0}) D_{mm}({\bf 0}) \notag\\
&+ \sum_{({\bf R}, n) \neq ({\bf 0}, m)} U_{m n} ({\bf R})D_{nn}({\bf 0})\notag\\
& - (1-\alpha) \sum_{({\bf R}, n) \neq ({\bf 0}, 0)} J_{m n}({\bf R}) D_{nn}({\bf R}),\\
&D_{mn}({\bf R}_{ij})= \sum_{\sigma} \left\langle c_{im \sigma}^{\dagger} c_{jn \sigma}\right\rangle_{\rm KS},
\end{align}
where $\alpha$ specifies how to treat the effects of the 
on-site Coulomb interactions and the exchange couplings.
In the standard calculation, we take $\alpha=1/2$.
Information regarding $D_{mn}({\bf R}_{ij})$ is outputted in \verb|zvo_dr.dat|. 

Corrections to the one-body terms other than the chemical potentials
can be defined as follows:
\begin{align}
&t_{mn}^{\rm DC}({\bf R}_{ij})= \frac{1}{2} J_{mn}({\bf R}_{ij}) (D_{nm}({\bf R}_{ji})\notag\\ 
&+2{\rm Re}[D_{nm}({\bf R}_{ji})])-\frac{1}{2} U_{mn}({\bf R}_{ij}) D_{nm}({\bf R}_{ji})\notag\\
&({\bf R}_{ij}, m) \neq ({\bf 0}, n).
\end{align}
However, these corrections originate from the Fock terms, and they are 
not fully included in the $ab$ $initio$ calculations.
Moreover, these corrections may greatly change the 
band structures in strong coupling regions. 
Therefore, we recommend not introducing these off-site corrections
in the actual calculations.
Nevertheless, 
$\HPhi$ can include these Fock-type corrections by a keyword \verb|Fock| because
it may be necessary for examining the effects of the off-site corrections.

\subsection{Calculation by $\HPhi$}
An example of an input file used with Standard mode of $\HPhi$
is shown as follows.
\begin{verbatim}
model = "Hubbard"
lattice = "Wannier90"
cutoff_length_t = 1
cutoff_length_u = 1
cutoff_length_j = 1
W = 2
L = 2
Height = 1
method = "CG"
2Sz = 0
nelec = 4
exct = 1
\end{verbatim}
By using \verb|lattice="Wannier90"|, 
$\HPhi$ reads the information of the effective Hamiltonians and
generates the input files.
Since the tractable systems sizes by $\HPhi$ is not large,
it is often necessary to omit the long-range hoppings and interactions,
whose length is larger than half of the linear dimensions of the system sizes.
To cut off the long-range part of the Hamiltonian, \verb|cutoff_length_t|,
\verb|cutoff_length_u|, and \verb|cutoff_length_j| are used.
If one wants to omit the small parameters in the Hamiltonians, 
one can specify the lower limit of the absolute values of the parameters by
\verb|cutoff_t|, \verb|cutoff_u|, and \verb|cutoff_j|.
For example, if $|t_{mn}(\vec{R_{ij}})|<$\verb|cutoff_t|, 
those hopping parameters are ignored.
A tutorial on the derivation and the analysis 
of the low-energy effective Hamiltonian for Sr$_{2}$CuO$_{3}$ can be found on the web page
of $\HPhi$~\cite{HPhi_tutorial}.

\section{Summary and Discussion}
\label{sec:summary}
To summarize, we first explained the installation of the current version of $\HPhi$
in Sec.~\ref{Sec:Installation}.
We also mentioned the submodule repository of Standard mode.
We also explained newly added functions/methods, such as 
full diagonalization using GPGPU and ScaLAPACK in Sec.~\ref{sec:fulldiag},
spectral calculations using the shifted Krylov method in Sec.~\ref{sec:spectrum},
the locally optimal block preconditioned conjugate gradient (LOBPCG) method
for obtaining several eigenvectors at once in Sec.~\ref{sec:CG},
real-time evolution in Sec.~\ref{sec:timeevo}, and
the canonical TPQ (cTPQ) method in Sec.~\ref{sec:cTPQ}.
In Sec.~\ref{sec:standard}, we explain the updates to
Standard mode such as the newly added keywords for methods (e.g., \verb|CG| for the LOBPCG method and
\verb|cTPQ| for the cTPQ method) and lattice structures (e.g. \verb|FCO| for the face-centered orthorhombic lattice and
\verb|Wannier90| for the Wannier90 format). 
In Sec.~\ref{sec:wan90}, we explained how to specify Hamiltonians
in the Wannier90 format. 
$\HPhi$ can perform the calculations for the $ab$ $initio$ low-energy effective Hamiltonians
obtained by {\tt{RESPACK}}, which outputs the information in the Wannier90 format,
as we explained in Sec. \ref{sec:respack}.

Here, we mention recently developed packages for exact diagonalization, 
such as QuSpin~\cite{quspin,Weinberg_SciPost2017,Weinberg_SciPost2019} and QS$^{3} $\cite{qs3,Ueda_CPC2022}. 
QuSpin is an open-source Python package for exact diagonalization, which can treat 
spin, fermionic, and bosonic systems. 
Besides ground-state calculations, quantum dynamics such as 
quench dynamics can also be performed using QuSpin.
QS$^{3}$ is a software package that focuses on performing 
exact diagonalization near fully polarized states for the spin-1/2 XXZ Heisenberg model. 
Since the Hilbert dimension near fully polarized states is small, it is possible to
have huge system sizes. For example, it is shown that
the exact diagonalization for a system size of $10^{3}=10\times10\times10$ 
with three down spins (the Hilbert dimension is about $1.6\times 10^{9}$) is possible using QS$^{3}$.
QS$^{3}$ also supports calculations of dynamical spin structure factors.
Compared to these software packages, 
the main advantage of $\HPhi$ is supporting massively parallel computation.
As we already showed in the previous paper~\cite{Kawamura_CPC2017}, the parallelization efficiency 
from 4096 cores to 32,768 cores reaches about 80\%. Thus, using massively parallel supercomputers, $\HPhi$ can perform not only ground-state calculations, but also finite-temperature calculations, excited-state calculations, and real-time evolution for a wide range of quantum lattice models.
Since each software package has its own advantage, 
the complementary use of several different 
packages is useful for efficiently completing research.

Lastly, we mention future extensions of $\HPhi$.
One is the implementation of the reduction of dimensions of the Hilbert space by using 
translational symmetry and point-group symmetry.
It is shown that the reduction of the dimensions enables us to perform
exact diagonalization calculations for up to 50 sites in spin-1/2 systems~\cite{Wietek_PRE2018}.
To reduce the {finite-size} effects, which is the main disadvantage of the
exact diagonalization method, the implementation of reduction by symmetry
is useful. Recently, an efficient algorithm~\cite{wallerberger} (trie-based ranking) is proposed
for finding bit patterns that satisfy a given condition.
Combination of the reduction of the Hilbert space and the recently proposed
efficient bit ranking algorithm can enable us to perform the
state-of-the-art exact diagonalization using $\HPhi$.
Another intriguing function is the finite-temperature spectrum calculations.
It is shown that the shifted Krylov method enables us to
perform exact finite-temperature spectrum calculations~\cite{Yamaji_arXiv2018}.
Implementing finite-temperature spectrum calculations is desirable for direct comparison with the experimental spectrum obtained at 
finite temperatures.
These further extensions will make $\HPhi$ a more useful
software package and will be reported in the near future.

\section*{Acknowledgement}
We wish to thank Takeo Hoshi, Tomohiro Sogabe, and Kazuma Nakamura 
for fruitful discussions. We also wish to thank {users of $\HPhi$} for their feedback and helpful contributions. 
TM thanks Tsuyoshi Okubo for useful discussions on the cTPQ method.
A part of the calculations was done using the Supercomputer Center, 
the Institute for Solid State Physics, the University of Tokyo.
KI, YM, KY, and TM were supported by Building of Consortia for 
the Development of Human Resources in Science and Technology, MEXT, Japan.
This study was also supported by Grants-in-Aid for Scientific Research
Nos. 19H01809, 20H00122, 20H01850, 23H01092, and 23H03818 from the Ministry of Education, Culture, Sports, Science and Technology, Japan.  
This work was also supported by the
National Natural Science Foundation of China (Grant No. 12150610462).
The development of $\HPhi$ was supported by ``Project for advancement of software usability in materials science" \cite{pasums} in the fiscal year 2015, 2016, 2017, and 2018.
The implementation of the full diagonalization 
using GPGPU was supported by the ``support service of program portability to General Purpose Graphics Processing Unit," which was operated by the Institute for Solid State Physics in the fiscal year 2017.





\bibliographystyle{elsarticle-num}
\bibliography{hphi}

\begin{thebibliography}{10}
\bibitem{CMA}
~We note that SpinCMA only supports the limited lattice geometry. For details,
  please refer the manual.
\end{thebibliography}







\end{document}